\documentclass[%
 reprint,
 amsmath,amssymb,
 aps,
]{revtex4-2}

\usepackage{graphicx}
\usepackage{dcolumn}
\usepackage{bm}
\usepackage{subfigure}
\usepackage{xcolor}
\usepackage{caption}
\usepackage{tabularray}
\UseTblrLibrary{booktabs}

\begin{document}


\title{\textbf{The USRA Feynman Quantum Academy:\\ If You Give a Student a Quantum Internship} 
}%

\author{Zoe Gonzalez Izquierdo}
 \email{Contact author: zgonzalez@usra.edu}
\affiliation{USRA Research Institute for Advanced Computer Science (RIACS), Mountain View, CA, 94043, USA}%
\author{Sophie Block}
\author{Besart Mujeci}
\author{P. Aaron Lott}
\affiliation{USRA Research Institute for Advanced Computer Science (RIACS), Mountain View, CA, 94043, USA}%
\author{Davide Venturelli}
\affiliation{USRA Research Institute for Advanced Computer Science (RIACS), Mountain View, CA, 94043, USA}%
\author{David Bell}
\affiliation{USRA Research Institute for Advanced Computer Science (RIACS), Mountain View, CA, 94043, USA}%


\begin{abstract}
In the rapidly expanding field of quantum computing, one key aspect to maintain ongoing progress is ensuring that early career scientists interested in the field get appropriate guidance and opportunity to advance their work, and in return that institutions and enterprises with a stake in quantum computing have access to a qualified pool of talent. Internship programs at the graduate level are the perfect vehicle to achieve this. In this paper, we review the trajectory of the USRA Feynman Quantum Academy Internship Program over the last 8 years, placing it in the context of the current push to prepare the quantum workforce of the future, and highlighting the caliber of the work it produced.
\end{abstract}

\maketitle

\section{\label{sec:intro} Introduction}

Internships---stepping stones between full-time student status and the first real-world job, providing the motivated student with a sandbox to test their newly acquired skills while keeping a safe level of hand holding; and the mentor with an opportunity to share wisdom, and perhaps provide the guidance and support they would have liked to receive at that stage of their careers.

When done right, internship programs embody a symbiotic relationship where the student benefits from the experience of the mentor and the institutional support, while they---and society as a whole---will harvest the fruits of investing in a promising young person for years to come, as the intern turns into a fully vested individual contributor and the future of the field is ensured. Any scientific endeavor benefits from successfully managing these programs.

Quantum technology is no different. In fact, given its relative recency when compared with other fields, it is easy to argue for the extra importance that the intern position holds here - as a rapidly growing field, degrees centered around it are not yet clearly defined and established, young folks choosing careers might not be familiar with its existence, much less know anyone in their circle that could guide them through its idiosyncrasies, and most of the information is found in scientific papers rather than more easily digestible textbooks. Creating and expanding quantum internship programs geared at different expertise levels is a sure way to help the burgeoning quantum scientist continue their career, someone coming from a different STEM background to pivot or add quantum knowledge to their skill set, or to steer a curious student into the start of their career in quantum.

Over the course of the past eight years during which the NASA Academic Mission Services (NAMS) contract was in place, the Universities Space Research Association (USRA)-led team tasked with its completion worked with NASA Ames Research Center to advance research topics such as air traffic management, autonomous systems, and quantum computing. The \emph{Feynman Quantum Academy} was created to promote the academic aspect of NAMS within the quantum task. Interns at USRA worked within the USRA-NASA Quantum Artificial Intelligence Laboratory (QuAIL) at NASA Ames Research Center~\cite{Rieffel_2019, Rieffel_2024}. QuAIL is the space agency's quantum computing research group, which, with Dr. Eleanor Rieffel at the helm, and a large portion of its technical workforce provided by USRA, produces research in a wide array of quantum information science areas, to help advance NASA missions. In addition to the intern projects supporting NASA, the Feynman Quantum Academy also led multiple intern projects in quantum computing through funding from AFRL, DOE, DARPA, DHS and NSF (see Table~\ref{tab:funding} for a breakdown of funding sources).

\begin{figure*}
    \centering
    \begin{subfigure}{}
        \includegraphics[width=0.9\linewidth]{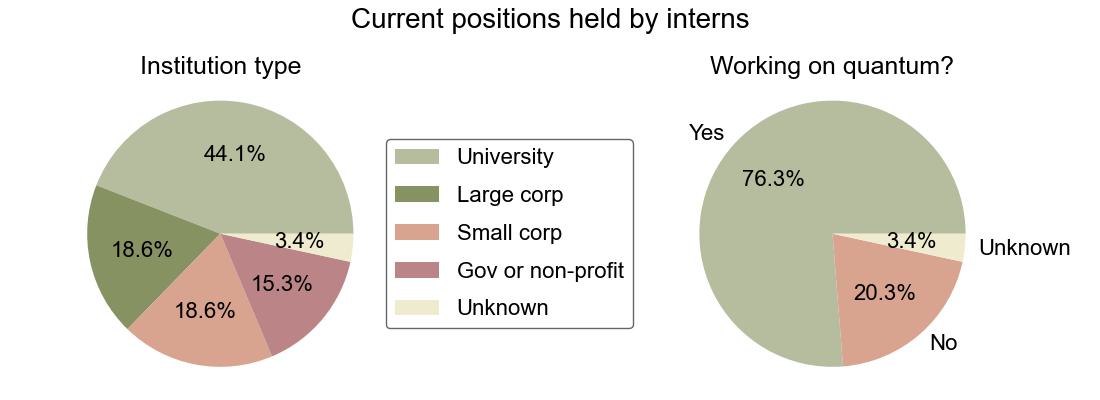}
        \caption{Breakdown of the current positions held by the Feynman Quantum Academy interns. A large portion work at Universities, as PhD students, postdocs, and associate and assistant professors. Large business, non-profit research institutions and startups or small business are roughly evenly split.  Remarkably, more than $3/4$ are still involved in quantum related work.}
        \label{fig:where_are_they_now}
    \end{subfigure}%
    \begin{subfigure}{}
        \includegraphics[width=0.8\linewidth]{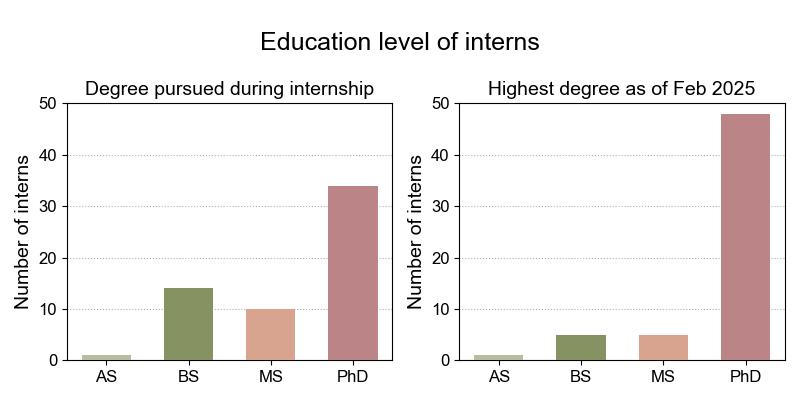}
        \caption{At the time of their internship, most students (34, or $~57\%$) were pursuing their PhD, with the next largest group being undergraduates (14 students, $~23\%$), 10 Masters students ($~17\%$) and a single Associates student ($<2\%$). Today, most of those who were in undergraduate or masters programs continued onto PhDs, with a total of 48 ($~81\%$) either having completed or being currently enrolled in a doctorate.}
        \label{fig:ed_level}
    \end{subfigure}%
\end{figure*}

\begin{table}
  \centering
  \begin{tblr}{
      colspec={ll},
      row{1}={font=\bfseries},
      row{even}={bg=gray!10},
    }
           \textbf{Funding source} & \textbf{Internships}   \\
    \toprule
    NASA & 49 \\
    NSF & 12 \\
    AFRL & 5 \\
    DARPA & 4 \\
    Fermilab & 2 \\
    DLR & 1 \\
    DHS & 1 \\
    \bottomrule
  \end{tblr}
\caption{Institutions or programs that have provided funding for the Feynman Academy, along with the number of internships each has funded (note that some internships were funded through more than one program).}
  \label{tab:funding}

\end{table}

\section{Background}
\subsection{The positive impact of internships}

The research supporting the benefits of internships is extensive and broad. Positive effects happen across the board, see e.g.~\cite{Binder_2015}. In this work, the effects of undergraduate internships on subsequent academic outcomes were explored through a longitudinal study covering a wide range of academic disciplines. Benefits remained robust across subjects, and while controlling for a number of factors in students' backgrounds, such as prior academic achievement.

Physics is no outlier when it comes to benefiting from this type of hands-on experience, as reported in~\cite{Lincoln_2019}. The authors emphasize the importance of going beyond the traditional classroom setting in order to foster appreciation for the subject. Summer internships can fill that gap, and provide students the opportunity to experience what a career in physics would look like, helping them strengthen their identity as scientists in the process.

An interesting case is the U.S. Compact Muon Solenoid (CMS) Program for Undergraduate Research Summer Experience (PURSUE)~\cite{Bose_2022,Banerjee_2024}. The U.S. CMS collaboration designed this novel internship program for undergraduates to gain experience working in High Energy Physics (HEP) and, more generally, help them develop the skills needed to succeed in STEM fields. The program has been running for several years, and participants fill out a survey after its conclusion. In this survey, all interns reported having improved their skills during the internship, and highly rated their motivation and enthusiasm to stay in a STEM career, with a majority stating their intention to pursue graduate studies in particle physics. They also found great value in the opportunity to network with and be mentored by scientists at Fermilab, and were able to use the connections established during the program to obtain meaningful letters of recommendation needed to apply to graduate school.

Much of the national internship apparatus is centered around the undergraduate stage. But graduate students, and in particular science, engineering and health (SEH) PhD students, should be allowed and encouraged to participate in internships while in graduate school, as the authors of~\cite{Murguia_2022} argue. Indeed, while SEH graduate programs are often very academics-focused, the current reality is that most SEH doctorates will work outside of academic institutions, and acquiring some familiarity with the environment where they will develop at least part of their career holds value for the student, as well as for the corporations and institutions that will later hire a better prepared candidate. Uncertainty about career prospects and lack of preparedness for satisfying careers in the real world are significant concerns of early career researchers. Off-campus internships during graduate school can alleviate these issues by developing and honing skills (in particular those that play a bigger role in non-academic setups), offering a new or wider perspective of potential applications of the student's field of research, and adding experience to their resume and connections to their career network.

Ref. \cite{McNeil_2017} agrees with the crucial role that internships play in setting up careers for physicists. While much more widely adopted by engineering programs, there are numerous reasons why they should be equally prevalent in physics ones. Of the 46\% of physics bachelors who entered the workforce after receiving their degrees (rather than immediately moving on to graduate school), 65\% are employed by the private sector, according to a survey of graduates from 2013 and 2014. Almost half of those who then pursue PhDs hold positions outside academia one year after receiving their degrees, and more of them move to private sector or government positions after completing a postdoctorate. 

Physics students leave their degrees typically well-versed in the core subjects of the curriculum, highly competent when it comes to solving advanced problems and conquering technical tasks. However, most would benefit from a deeper knowledge of the nontechnical aspects of science that come into play when being employed in non-academic scientific and technical roles, such as computational tools used in industry, communication skills, and project execution and documentation in a business environment. Off-campus internships are a means to honing these and other skills required to succeed in the workplace, providing exposure to the connection among physics content, applications, and innovation. 

This is not a new issue, as evidenced by this letter from 1995~\cite{Withheld_1995}, lamenting that the standard university experience was not sufficient preparation for physicist roles in industry. The number of physics students that end up in non-academic positions has only continued to grow over the last few decades. Now that technology has permeated all aspects of life and the Artificial Intelligence (AI) revolution is solidly entrenched in the global economy, STEM majors have a wider than ever range of options to choose from, and they would do well to familiarize themselves with the types or careers available in government and the private sector. 

\subsection{Preparing the quantum workforce of the future}

2025 has been declared by the United Nations as the International Year of Quantum Science and Technology (IYQ), which will ``be observed through activities at all levels aimed at increasing public awareness of the importance of quantum science and applications.''~\cite{UN2025} The expansion of the quantum computing field from the purely academic ambit into the realm of businesses and government institutions is well underfoot, and will require a workforce equipped with the appropriate technical expertise, that is also ready to succeed in industry. Institutional investment and involvement will have a huge impact on the development of the new generation of quantum scientists, to the benefit of society as a whole.

Governments across the world have taken notice. In 2016 the US federal government identified quantum workforce development as a priority, citing that “academic and industry representatives identify discipline-specific education as insufficient for continued progress in quantum information science.”~\cite{advancing_qis}. Then the National Quantum Iniciative (NQI) Act was passed in 2018, which included the goal to “expand the number of researchers, educators, and students with training in quantum information science and technology to develop a workforce pipeline.”~\cite{NQIA}. How exactly to accomplish this has been a topic of research over the past few years, and through a large body of work involving all stakeholders, some general consensus and trends emerge.

One thing is abundantly clear: quantum information science and technology (QIST) (or quantum information science and engineering (QISE)) brings together a rare combination of academic, industrial and government interest, support and financial investment, along with the talent and passion of its current workforce. We must make it possible for all sides to work together and capitalize on this momentum. The literature broadly agrees that the current state of quantum education is not fit to tackle the needs of the flourishing quantum industry. The Quantum Economic Development Consortium (QED-C) was established in the US with support from the National Institute of Standards and Technology (NIST) as part of the Federal strategy for advancing quantum information science and as called for by the NQI Act. The QED-C is a consortium of stakeholders with the goal to support the quantum industry, whose participants are working together to identify gaps in technology, standards, and workforce and collaboratively address them.

Shortly after its formation, many of the U.S. companies that signed letters of intent with the QED-C participated in a study of the quantum industry~\cite{Fox_2020} to better understand what skills are most needed and which are currently difficult to find when hiring, and use those findings to offer educational guidance. Based on companies' responses, a categorization of activities was proposed, identifying five main types of technical careers available within the quantum industry (engineers, experimental scientists, theorists, technicians and application researchers), and laying out their different skillsets, backgrounds and responsibilities.

Most interviewees reported that we are still at a point where employees usually need a Ph.D. (typically physics, or related disciplines) to be useful in the field. However, as the industry develops, and the technology transitions from research to product, a larger proportion of engineers and technicians will be needed. In preparation for this upcoming shift, a key request for higher-education institutions is a one- or two-semester course in order to increase quantum awareness, covering basic aspects of quantum information and geared to engineering backgrounds. They also noted that the (typically) physics PhDs working in the field, while highly capable in their technical area of expertise, have room for improvement when it comes to other necessary skills, such as software development in a collaborative environment, general team-working skills, engineering and system design, and a better understanding of how businesses work.

Other work agrees with this assessment. Education and workforce training in QIST exists primarily at the graduate and postdoctoral levels in the US, and will need to be expanded to the undergraduate level to meet the anticipated quantum workforce needs~\cite{Perron_2021}. To that end, the Quantum Undergraduate Education and Scientific Training (QUEST) workshop was held in 2021, bringing together faculty from predominantly undergraduate institutions (PUIs) to learn the state of undergraduate QIST education, identify challenges associated with implementing QIST curriculum at their institutions, and to develop solutions to these challenges. They concluded that academic research efforts in QIST must move beyond the larger PhD granting institutions that currently produce the majority of this work, and into PUIs which are ideally suited to reach the larger number of students that will be needed as the quantum industry expands.

To effectively train the emerging quantum workforce, universities and colleges require knowledge of the type of jobs available for their students, and of which skills and degrees are most relevant for those new jobs, and organizations with the right capabilities must work on developing the appropriate curricula and training methods. A multitude of efforts in both of these directions have taken place over the past few years. The lowest hanging fruit are master's programs in QISE, as these will be closest in scope and level to existing PhD programs, and will likely not require developing dedicated courses from scratch, but rather working out how to tailor and maximize a year or two worth of education to best prepare students for their future jobs, and then selecting, editing and organizing a set of relevant graduate-level courses (and potentially other training materials). This type of master's started being offered by universities a few years ago and has kept expanding ever since. 

In 2019, a symposium of 50 US and European QISE experts from both industry and academia discussed post-secondary education with an emphasis on master’s programs~\cite{Aiello_2021}, by studying the status of eighteen such programs. They expressed support for the establishment of a comprehensive strategic plan for quantum education and workforce development as a means to make the most of the ongoing substantial investments being made in QISE, and consider that the rapid pace of innovation in the field makes it challenging to develop standardized classroom- and lab-based curricula, necessitating the recruitment of both subject-matter experts and experts in building interdisciplinary programs outside of QISE to tackle the challenge.

From the insights gained after conducting a survey of companies in the quantum industry, \cite{Hughes_2022} makes some similar recommendations. The authors consider that the vast majority of new jobs will not require many quantum-specific skills, so educators developing new master’s programs should strike a balance between quantum specific courses and more general STEM courses. And that educational approaches other than full degrees are more likely to provide utility to the quantum industry, so institutions should consider offering one or two broad quantum courses for the population that will need some level of quantum knowledge, such as students aiming to fill business roles in the quantum industry. They also note that there is greater consensus about which skills--rather than degrees--are important, and so non-coursework experiences and training based on those skills, such as summer internships, should be prioritized. 

There is a strong focus on designing and implementing these educational paths to address the expansion of the quantum workforce. In 2022, the US Government published the Quantum Information and Technology Workforce Development Plan, and one of their key strategies is broadening student participation in post-secondary education--with a focus on quantum skills--at all levels, including associate’s, bachelor’s, master’s, and professional training programs, to address workforce gaps. At the bachelor’s level, the rapidly growing QISE industry will require both quantum-aware and quantum-proficient engineers. Ref. \cite{Asfaw_2022} presents a comprehensive road map for building a quantum engineering education program to satisfy both needs. The authors point out this is a pressing issue as a gap is left between, on the one hand, excitement generated by popular media leading to a push to introduce QISE in secondary school, and on the other hand, quantum-related graduate programs focused mainly on PhDs, with some MS programs as well. This can be addressed in the near term and will likely have a substantial positive impact on the quantum workforce.

The multiple points of disconnect between academia and industry are a frequent theme in the literature, and we must understand the current quantum ecosystem’s competencies and education needs, as well as any existing educational initiatives to see the full picture. Ref. \cite{Maninder_2022} presents an overview of this whole ecosystem, focusing on where to find all the learning and training resources to start a career in quantum technologies, and what activities and resources are best suited to train that workforce. One of their key takeaways is that academic degrees alone cannot provide industry desirable job-ready skills, and industry experience and up-skilling should be utilized to tackle many of the workforce needs.

It is clear that a multi-prong approach spanning a range of initiatives will be needed~\cite{Raghunathan_2023}, including a combination of conventional coursework, project-based learning and institutional partnerships. The short-term demands of this growing industry for talent that understands the core concepts of QIST, can be addressed by incorporating quantum skills and training through different sources and activities, including online courses, conferences, workshops, hackathons, games, and community-building forums, which can be part of the initiatives for retraining and up-skilling the existing workforce. Master’s and PhD programs are already in place and continue to be developed and refined to address the workforce at a longer term. For a much longer term, the education of current high school and primary school students will become relevant. Indeed, the QIST Workforce Development National Strategic Plan also recommends fostering precollege interest in QIST as students develop career aspirations. There is a disconnect between the need for precollege and undergraduate students to be aware of and recognize QIST as a viable career option, and the current education system in which QIST topics are usually not covered until the advanced undergraduate level. \cite{Darienzo_2024} explored the outcomes from a one-week, 25-hour summer program for US high school students to become acquainted with the industry and its potential job opportunities. While the sample group was small, it appeared that all of the participating students, which were in grades 10-12 and had varying levels of STEM backgrounds, improved their disciplinary knowledge and awareness of vocational roles in QIST.
 
 Some programs have already been implemented at a more specialized, technical level. In 2019, the National Science Foundation introduced the Quantum Leap Challenge Institute (QLCI)~\cite{qlci} program as part of its Quantum Leap Big Idea. Three QLCIs were funded in the inaugural round of awards in 2020, with two more funded the following year. One example is Q-SEnSE~\cite{qsense}, drawing from the strong tradition in quantum sensing and metrology research at CU Boulder, its lead institution, as well as at ten other academic institutions. They promote a specialist education in an academic setting as well as content aimed at attaining only a basic quantum understanding, but based on a strong vocational foundation, as it is their belief that these are both needs that quantum education should fulfill~\cite{Bennett_2024}. The EdQuantum project~\cite{edquantum}, funded through the National Science Foundation (NSF) Advanced Technological Education (ATE) program, is another effort that aims to fulfill the upcoming need for highly skilled quantum technicians able to support the commercialization of the new products and inventions. This is an area which is still coming up short, despite general considerable investment in quantum research~\cite{Hasanovic_2022}. EdQuantum posits that one of the reasons for the current gap between the advances in quantum science and their potential industrial applications is due to this lack of technical workers skilled in quantum technologies, and that photonics technicians are in a favorable position of already possessing many of the required skills, so that a relatively short training can get them fully ready as quantum technicians.
CU Boulder offers another option for expanding the quantum workforce via a senior capstone course called “Quantum Forge”, with a goal for students to understand what comprises the quantum industry and become confident that they could participate in it if desired. During this course, students partner with a company in the quantum industry to work on an authentic project. Insights from the first cohort were gathered and published in~\cite{Oliver_2024}, with the main takeaway being that the most important job preparation for students is acquiring relevant experience in the area of interest. The participants did not know very much about the quantum industry before the course, but they were able to become familiar with the available jobs and left with a more clear idea of whether this industry was a good fit for them.

Initiatives such as the NQI Act in the US, or the Quantum Flagship in Europe, both established in 2018, intend to bring their respective geographic areas up to speed with quantum technologies. The European Quantum Readiness Center (EQRC), funded by the European Commission through the Quantum Flagship initiative, will help ensure that companies and institutions throughout Europe are at the forefront of quantum technology, and the European Quantum Industry Consortium (QuIC) has created a strategic industrial road map to shape the education and training of the European quantum-ready workforce. 

The findings from the European side echo the same sentiment as those from the US, highlighting that now is the time to start training the future quantum-literate workforce. These efforts are driven by the Quantum Flagship and the corresponding Coordination and Support Action for Quantum Technology Education (QTEdu CSA)~\cite{qtedu}. Within this context, an initial Delphi study~\cite{Gerke_2022} was carried out to scope out knowledge and skill requirements for future quantum professionals, by enlisting the help of experts' opinions from research, industry and teaching. A salient point of the study was that this workforce will be made up of people who work with quantum technologies, but who do not need as strong a quantum physics background as physicists. Knowledge and competence areas required for this group of people can be called “Quantum Awareness”, and will differ from those that the typical quantum mechanics courses for physicists address.  

These are the first steps towards an European certification scheme to standardize industry training and educational efforts, using the European Competence Framework for Quantum Technologies as the common language, and being compiled within the QTEdu project as well as the Quantum Flagship’s new Coordination and Support Action (QUCATS)~\cite{qucats}. An initial report~\cite{Greinert_2023} and follow-up~\cite{Greinert_2023_2} lay out the progress in collecting and identifying these competences.

Two additional EU-funded projects are the DigiQ project~\cite{digiq} for developing master’s programs in quantum technology across Europe, and the QTIndu project~\cite{qtindu} for upskilling the existing industry workforce to be quantum-ready. A follow-up to the previous two reports~\cite{Greinert_2024} provides further insights into industry needs and how to facilitate the development of more effective and practical training and education strategies, through interviews with relevant people in industry.

\section{The Feynman Quantum Academy}

Considering all of the above, the need for internships in quantum computing is undeniable. The Feynman Quantum Academy brings interns the opportunity to work within the USRA-NASA Quantum Artificial Intelligence Laboratory (QuAIL) at NASA Ames Research Center, bridging the gap between educational programs and industry applications to address the critical demand for a workforce equipped for quantum technology roles. Working alongside industry collaborators, the Academy provides a curriculum that serves both present and future industry demands. It strikes a balance between theoretical understanding and applied skills, spanning from basic quantum mechanics to sophisticated computing methods. The program spotlights the competencies that are most sought after, promoting an environment of teamwork and inventive thought.

Students are typically  enrolled in a PhD program during their time as an intern (although particularly strong candidates that are pursuing an MSc or BS can be considered). Most have at least some previous knowledge of quantum computing, and their degrees are in a related field such as Physics, Mathematics, Electrical Engineering or Computer Science, but some candidates might still do well without quantum computing experience if they have expertise in other technical skills that are combined with quantum (e.g. machine learning, optimization, software engineering etc.).

The internships usually last 12 to 24 weeks, occasionally getting extended beyond that range. Applications are open year-round, with rolling acceptances, and summer being the most popular time of the year (when most University programs have more flexibility).  Internships during the academic semester are often part-time, to enable a balance of coursework along with research. Until 2019, the program was fully on-site at the NASA Ames Research Center in Mountain View, California. Since 2020, it has mostly remained remote, although students that are local to the area, or able to relocate or visit, are encouraged to participate in person. Living expenses during the program are covered through a stipend.

USRA research staff members periodically meet to review internship applications and identify candidates whose experience and interests fit with an existing or upcoming project. After the candidate is successfully interviewed and a start date agreed upon, they join the group and participate as any other member. The student works in close collaboration with one of the research scientists (sometimes two), who is their mentor within the group, typically meeting weekly to discuss their ongoing work and receive guidance as needed. The students also participate in group meetings, seminars and opportunities for networking and collaborating with other quantum scientists, both from the QuAIL group and from other academic institutions, national laboratories or corporations. Finally, the student will produce a code and/or publication from the work performed during the internship. Oftentimes, they are able to include this work into their doctoral thesis. Some highly productive interns manage to output several papers.

Generally, these experential learning opportunities provide access to quantum processors through programs led by NASA or USRA in collaboration with hardware vendors, such as Rigetti and IBM, as well as high-performance computing hardware through the NASA Advanced Supercomputer (NAS) and/or the NSF National Research Platform (NRP). This mix of QPU, CPU, GPU and FPGA access allows for benchmarking, simulation and algorithm development and analysis. The internships are often motivated by an industrial and government application relevant to the funding program involving problems from aeronautics or defense, that commonly inform requirements for particular optimization, constraint satisfaction, machine learning or simulation parameters of interest. The interns are also embedded within a team of researchers focused on assessing and advancing some aspect of quantum technology. Access to these three types of resources provides interns a context and basis of support to fill common gaps encountered in a strictly academic enviroment. This blend of resources also provides academic professors, students, industry and government researchers a means for collaborating, networking and energizing the stakeholders to address problems of mutual interest with unique perspectives. While not all interns and projects leverage the full set of resources, through group meetings and journal club presentations interns have opportunities to observe how classical high-performance computing, machine learning, optimization and numerical simulation can serve serve as current state-of-the-art baselines driving applications of relevance to government and industry and as targets for integrating with future quantum technology to improve upon state-of-the-art.

Just like those of the QuAIL team itself, the research and publications of the Feynman Quantum Academy students span a wide range of topics in quantum computation and adjacent fields. The review articles~\cite{Biswas_2017,Rieffel_2019,Rieffel_2024} provide a thorough overview of the work produced by QuAIL over the last decade.

\begin{figure*}[ht] 
    \centering
    \includegraphics[width=0.75\textwidth]{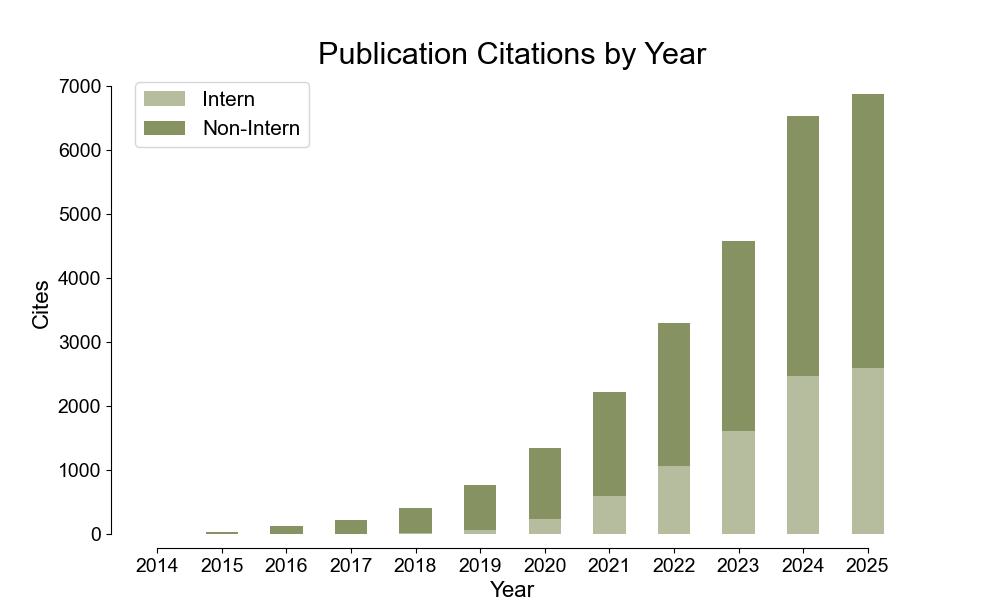} 
    \caption{Citations by year of all publications (2014-April 2025) of USRA, highlighting the split between papers in which one or more of the authors were interns vs. papers without intern participation.}
    \label{fig:citations}
\end{figure*}

\begin{figure*}
        \includegraphics[width=\textwidth]{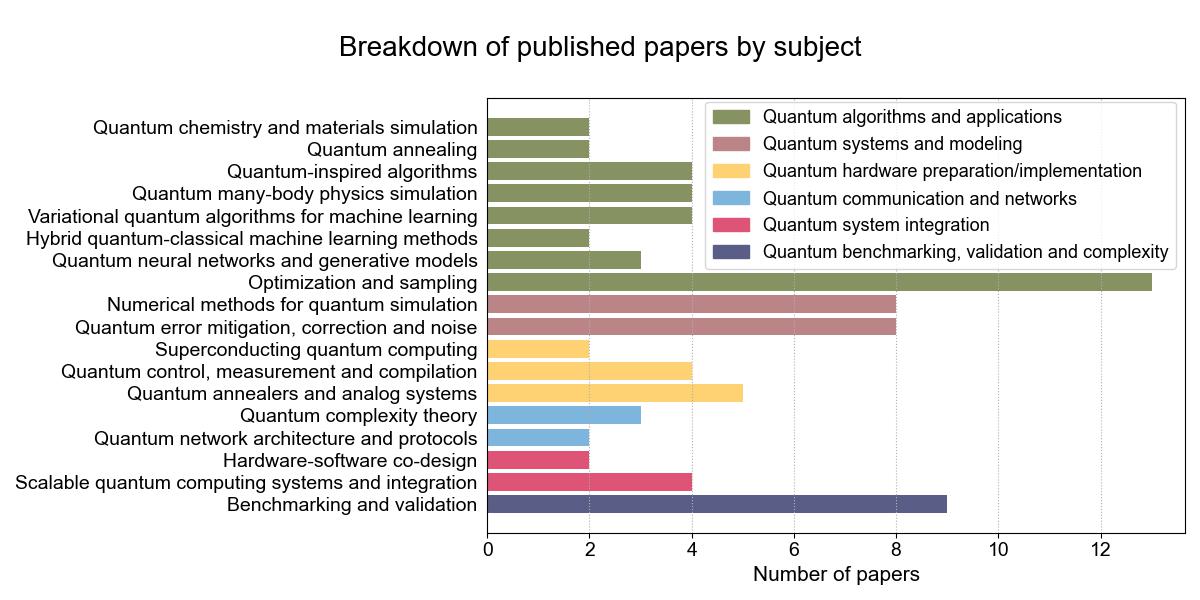}
        \caption{Each paper where interns collaborated is categorized by main topic (represented by color), then
by sub-category (written on the y axis). Each paper can be assigned to more that one subject.}
        \label{fig:topics}
\end{figure*}

While the volume of publications by interns alone is noteworthy, even more remarkable is the impact of these publications on the quantum community. Fig.~\ref{fig:citations} compares the citations garnered by papers with Feynman Quantum Academy intern collaboration to those without such collaboration over the past decade, illustrating that papers involving interns have a comparable impact (as measured by citations) to papers without intern involvement. So not only are interns actively publishing, but their work is also receiving recognition and attention comparable to other researchers in the field. A distribution of paper topics is presented in Fig.~\ref{fig:topics}. For a thorough review of the work organized by subject we refer the reader to Appendix~\ref{research}.

\section{Related programs}

There are several other high quality programs that offer similar experiences in the quantum science and technology space. One such example is the Quantum Computing Summer School at Los Alamos National Lab (LANL)~\cite{los_alamos}. Established in 2018, and having run every summer since then, it is similar to the Feynman Quantum Academy in scope, bringing the opportunity to conduct research in quantum computing while being mentored by experts in the field to graduate (and motivated undergraduate) students.

In contrast to the Feynman Quantum Academy, the program at LANL is designed as a more traditional summer school, with a predetermined duration and schedule, rather than the more ad hoc setup at QuAIL. It lasts a total of 10 weeks, with the next occurrence starting June 2nd, 2025, and applications closing on January 19th, 2025. The first two weeks are exclusively dedicated to lectures given by different subject matter experts. Then each student is paired with an LANL mentor to guide them through their research project over the remainder 8 weeks, which includes hands-on programming of one of the available quantum computers. Students are expected to come from a STEM background, and be either at the later stages of an undergraduate degree or the early ones of a graduate degree. Interestingly, we observed that multiple Feynman Academy students participated earlier or subsequently to the LANL internship program.

Another related program is the SQMS Quantum Undergraduate Internship at Fermi National Accelerator Laboratory (Fermilab)~\cite{sqms}. The SQMS (Superconducting Quantum Materials and Systems Center) is one of five research centers funded by the U.S. Department of Energy to advance quantum computers and sensors. Led by Fermilab, and conceived as part of a large national initiative, it fosters collaboration among hundreds of experts from a range of academic and industrial backgrounds, with work spanning all aspects of quantum computation and technology, from hardware to algorithms, sensing or the fundamental physics of quantum devices.

Sponsored by the Department of Energy Office of Science, the SQMS Quantum Undergraduate Internship places undergraduate sophomore and junior physics and engineering majors in a paid 10-week summer internship program at the SQMS Center. Students gain hands-on experience through access to state-of-the-art facilities under the mentorship of SQMS experts across the Center’s 24 partner institutions.

This paid internship program is open to undergraduate students in the second or third year of their physics or engineering degree, occurs over a 10-week period during the summer, and can take place at the Fermilab campus or at one of its many partner institutions.
Students have the opportunity to work on campus at Fermilab or at one of the center’s partner institutions, including NASA Ames Research Center.

In Canada, there is a large program funded by Mitacs, a non-profit national research organization that partners with academic, government and business institutions with the goal of advancing innovation across a variety of fields. They offer a range of programs aimed at different educational levels, with their largest one being Accelerate, which has been providing internship opportunities to graduate students and postdoctoral fellows since 2004. Mitacs partners with over 65 Canadian institutions in the quantum and AI fields, including the Institute for Quantum Computing (IQC) at the Univeristy of Waterloo, D-Wave Systems, 1QBit, Xanadu, etc. Companies and organizations can sign up to be matched with top research talent that fits the specific needs of their research project, and Mitacs covers up to 50\% of the cost. Students at any college level can apply, as well as postdocs that are within the first five years post-graduation. Internships start at 4 months, some lasting up to several years, and there is a large amount of inter-institutional collaboration. 

\section{Conclusion}

The Feynman Quantum Academy program has supported 60 student internships between 2016 and 2024, with several interns having participated in multiple projects that lasted more than one year (so in some cases the totals in prior sections add up to over 60). While the program has generally focused on PhD-level research projects, it has supported students at a range of educational levels; from some who have just exited high-school to others enrolled in local community colleges. These experential learning opportunities provide access to quantum and high-performance computing hardware, industrial and government applications and a team of researchers focused on assessing and advancing emerging quantum technology, filling common gaps left by a strictly academic environment. This blend of resources also provides academic professors, students, industry and government a means for collaborating, networking and energizing stakeholders to address problems of mutual interest with unique perspectives. Students involved through the program have generally continued to work in the quantum information sciences (over 75$\%$ remain in the quantum industry in some capacity at this time), either by continuing their education, or through obtaining academic or industrial research positions in promising startups or fortune 500 companies such as Google, Amazon, IBM, JP Morgan Bank and Boeing. 

The Feynman Quantum Academy has demonstrated a proven model for experiential learning in quantum computing, equipping students with the skills and knowledge necessary to drive innovation in quantum algorithms, hybrid AI-quantum systems, and real-world applications of these emerging technologies. As the demand for quantum-literate scientists and engineers continues to grow, expanding access to hands-on training, mentorship, and interdisciplinary research opportunities will be critical in shaping the next generation of quantum leaders.

We invite academic institutions, industry partners, and government agencies to collaborate in scaling this model, ensuring that the global quantum ecosystem remains inclusive and at the cutting edge of technological advancements. By investing in quantum education and workforce development, we can collectively accelerate progress toward a future where quantum computing and AI-driven innovations address some of the most complex scientific and societal challenges.

To join this effort, we encourage prospective students, researchers, and partners to engage with the Feynman Quantum Academy, contribute to its mission, and help shape the next wave of quantum breakthroughs.


\acknowledgments

This review of the Feynman Quantum Academy was funded by NSF CCF 1918549.

\bibliography{references}

\begin{thebibliography}{79}%
\makeatletter
\providecommand \@ifxundefined [1]{%
 \@ifx{#1\undefined}
}%
\providecommand \@ifnum [1]{%
 \ifnum #1\expandafter \@firstoftwo
 \else \expandafter \@secondoftwo
 \fi
}%
\providecommand \@ifx [1]{%
 \ifx #1\expandafter \@firstoftwo
 \else \expandafter \@secondoftwo
 \fi
}%
\providecommand \natexlab [1]{#1}%
\providecommand \enquote  [1]{``#1''}%
\providecommand \bibnamefont  [1]{#1}%
\providecommand \bibfnamefont [1]{#1}%
\providecommand \citenamefont [1]{#1}%
\providecommand \href@noop [0]{\@secondoftwo}%
\providecommand \href [0]{\begingroup \@sanitize@url \@href}%
\providecommand \@href[1]{\@@startlink{#1}\@@href}%
\providecommand \@@href[1]{\endgroup#1\@@endlink}%
\providecommand \@sanitize@url [0]{\catcode `\\12\catcode `\$12\catcode `\&12\catcode `\#12\catcode `\^12\catcode `\_12\catcode `\%12\relax}%
\providecommand \@@startlink[1]{}%
\providecommand \@@endlink[0]{}%
\providecommand \url  [0]{\begingroup\@sanitize@url \@url }%
\providecommand \@url [1]{\endgroup\@href {#1}{\urlprefix }}%
\providecommand \urlprefix  [0]{URL }%
\providecommand \Eprint [0]{\href }%
\providecommand \doibase [0]{https://doi.org/}%
\providecommand \selectlanguage [0]{\@gobble}%
\providecommand \bibinfo  [0]{\@secondoftwo}%
\providecommand \bibfield  [0]{\@secondoftwo}%
\providecommand \translation [1]{[#1]}%
\providecommand \BibitemOpen [0]{}%
\providecommand \bibitemStop [0]{}%
\providecommand \bibitemNoStop [0]{.\EOS\space}%
\providecommand \EOS [0]{\spacefactor3000\relax}%
\providecommand \BibitemShut  [1]{\csname bibitem#1\endcsname}%
\let\auto@bib@innerbib\@empty
\bibitem [{\citenamefont {{Rieffel}}\ \emph {et~al.}(2019)\citenamefont {{Rieffel}}, \citenamefont {{Hadfield}}, \citenamefont {{Hogg}}, \citenamefont {{Mandr{\`a}}}, \citenamefont {{Marshall}}, \citenamefont {{Mossi}}, \citenamefont {{O'Gorman}}, \citenamefont {{Plamadeala}}, \citenamefont {{Tubman}}, \citenamefont {{Venturelli}}, \citenamefont {{Vinci}}, \citenamefont {{Wang}}, \citenamefont {{Wilson}}, \citenamefont {{Wudarski}},\ and\ \citenamefont {{Biswas}}}]{Rieffel_2019}%
  \BibitemOpen
  \bibfield  {author} {\bibinfo {author} {\bibfnamefont {E.~G.}\ \bibnamefont {{Rieffel}}}, \bibinfo {author} {\bibfnamefont {S.}~\bibnamefont {{Hadfield}}}, \bibinfo {author} {\bibfnamefont {T.}~\bibnamefont {{Hogg}}}, \bibinfo {author} {\bibfnamefont {S.}~\bibnamefont {{Mandr{\`a}}}}, \bibinfo {author} {\bibfnamefont {J.}~\bibnamefont {{Marshall}}}, \bibinfo {author} {\bibfnamefont {G.}~\bibnamefont {{Mossi}}}, \bibinfo {author} {\bibfnamefont {B.}~\bibnamefont {{O'Gorman}}}, \bibinfo {author} {\bibfnamefont {E.}~\bibnamefont {{Plamadeala}}}, \bibinfo {author} {\bibfnamefont {N.~M.}\ \bibnamefont {{Tubman}}}, \bibinfo {author} {\bibfnamefont {D.}~\bibnamefont {{Venturelli}}}, \bibinfo {author} {\bibfnamefont {W.}~\bibnamefont {{Vinci}}}, \bibinfo {author} {\bibfnamefont {Z.}~\bibnamefont {{Wang}}}, \bibinfo {author} {\bibfnamefont {M.}~\bibnamefont {{Wilson}}}, \bibinfo {author} {\bibfnamefont {F.}~\bibnamefont {{Wudarski}}},\ and\ \bibinfo {author} {\bibfnamefont {R.}~\bibnamefont {{Biswas}}},\ }\bibfield
  {title} {\bibinfo {title} {{From Ans{\"a}tze to Z-gates: a NASA View of Quantum Computing}},\ }\bibfield  {journal} {\bibinfo  {journal} {IOS Press}\ }\href {https://doi.org/10.3233/APC190010} {10.3233/APC190010} (\bibinfo {year} {2019})\BibitemShut {NoStop}%
\bibitem [{\citenamefont {Rieffel}\ \emph {et~al.}(2024)\citenamefont {Rieffel}, \citenamefont {Asanjan}, \citenamefont {Alam}, \citenamefont {Anand}, \citenamefont {{Bernal Neira}}, \citenamefont {Block}, \citenamefont {Brady}, \citenamefont {Cotton}, \citenamefont {{Gonzalez Izquierdo}}, \citenamefont {Grabbe}, \citenamefont {Gustafson}, \citenamefont {Hadfield}, \citenamefont {Lott}, \citenamefont {Maciejewski}, \citenamefont {Mandrà}, \citenamefont {Marshall}, \citenamefont {Mossi}, \citenamefont {Bauza}, \citenamefont {Saied}, \citenamefont {Suri}, \citenamefont {Venturelli}, \citenamefont {Wang},\ and\ \citenamefont {Biswas}}]{Rieffel_2024}%
  \BibitemOpen
  \bibfield  {author} {\bibinfo {author} {\bibfnamefont {E.~G.}\ \bibnamefont {Rieffel}}, \bibinfo {author} {\bibfnamefont {A.~A.}\ \bibnamefont {Asanjan}}, \bibinfo {author} {\bibfnamefont {M.~S.}\ \bibnamefont {Alam}}, \bibinfo {author} {\bibfnamefont {N.}~\bibnamefont {Anand}}, \bibinfo {author} {\bibfnamefont {D.~E.}\ \bibnamefont {{Bernal Neira}}}, \bibinfo {author} {\bibfnamefont {S.}~\bibnamefont {Block}}, \bibinfo {author} {\bibfnamefont {L.~T.}\ \bibnamefont {Brady}}, \bibinfo {author} {\bibfnamefont {S.}~\bibnamefont {Cotton}}, \bibinfo {author} {\bibfnamefont {Z.}~\bibnamefont {{Gonzalez Izquierdo}}}, \bibinfo {author} {\bibfnamefont {S.}~\bibnamefont {Grabbe}}, \bibinfo {author} {\bibfnamefont {E.}~\bibnamefont {Gustafson}}, \bibinfo {author} {\bibfnamefont {S.}~\bibnamefont {Hadfield}}, \bibinfo {author} {\bibfnamefont {P.~A.}\ \bibnamefont {Lott}}, \bibinfo {author} {\bibfnamefont {F.~B.}\ \bibnamefont {Maciejewski}}, \bibinfo {author} {\bibfnamefont {S.}~\bibnamefont {Mandrà}}, \bibinfo
  {author} {\bibfnamefont {J.}~\bibnamefont {Marshall}}, \bibinfo {author} {\bibfnamefont {G.}~\bibnamefont {Mossi}}, \bibinfo {author} {\bibfnamefont {H.~M.}\ \bibnamefont {Bauza}}, \bibinfo {author} {\bibfnamefont {J.}~\bibnamefont {Saied}}, \bibinfo {author} {\bibfnamefont {N.}~\bibnamefont {Suri}}, \bibinfo {author} {\bibfnamefont {D.}~\bibnamefont {Venturelli}}, \bibinfo {author} {\bibfnamefont {Z.}~\bibnamefont {Wang}},\ and\ \bibinfo {author} {\bibfnamefont {R.}~\bibnamefont {Biswas}},\ }\bibfield  {title} {\bibinfo {title} {Assessing and advancing the potential of quantum computing: A nasa case study},\ }\href {https://doi.org/https://doi.org/10.1016/j.future.2024.06.012} {\bibfield  {journal} {\bibinfo  {journal} {Future Generation Computer Systems}\ }\textbf {\bibinfo {volume} {160}},\ \bibinfo {pages} {598} (\bibinfo {year} {2024})}\BibitemShut {NoStop}%
\bibitem [{\citenamefont {Binder}\ \emph {et~al.}(2015)\citenamefont {Binder}, \citenamefont {Baguley}, \citenamefont {Crook},\ and\ \citenamefont {Miller}}]{Binder_2015}%
  \BibitemOpen
  \bibfield  {author} {\bibinfo {author} {\bibfnamefont {J.~F.}\ \bibnamefont {Binder}}, \bibinfo {author} {\bibfnamefont {T.}~\bibnamefont {Baguley}}, \bibinfo {author} {\bibfnamefont {C.}~\bibnamefont {Crook}},\ and\ \bibinfo {author} {\bibfnamefont {F.}~\bibnamefont {Miller}},\ }\bibfield  {title} {\bibinfo {title} {The academic value of internships: Benefits across disciplines and student backgrounds},\ }\href {https://doi.org/https://doi.org/10.1016/j.cedpsych.2014.12.001} {\bibfield  {journal} {\bibinfo  {journal} {Contemporary Educational Psychology}\ }\textbf {\bibinfo {volume} {41}},\ \bibinfo {pages} {73} (\bibinfo {year} {2015})}\BibitemShut {NoStop}%
\bibitem [{\citenamefont {Lincoln}\ \emph {et~al.}(2019)\citenamefont {Lincoln}, \citenamefont {Pasero},\ and\ \citenamefont {Thompson}}]{Lincoln_2019}%
  \BibitemOpen
  \bibfield  {author} {\bibinfo {author} {\bibfnamefont {D.}~\bibnamefont {Lincoln}}, \bibinfo {author} {\bibfnamefont {S.}~\bibnamefont {Pasero}},\ and\ \bibinfo {author} {\bibfnamefont {R.}~\bibnamefont {Thompson}},\ }\bibfield  {title} {\bibinfo {title} {Making future researchers: Internship opportunities for physics students},\ }\href {https://doi.org/10.1119/1.5135801} {\bibfield  {journal} {\bibinfo  {journal} {The Physics Teacher}\ }\textbf {\bibinfo {volume} {57}},\ \bibinfo {pages} {642} (\bibinfo {year} {2019})},\ \Eprint {https://arxiv.org/abs/https://pubs.aip.org/aapt/pte/article-pdf/57/9/642/9871783/642\_1\_online.pdf} {https://pubs.aip.org/aapt/pte/article-pdf/57/9/642/9871783/642\_1\_online.pdf} \BibitemShut {NoStop}%
\bibitem [{\citenamefont {Bose}\ \emph {et~al.}(2022)\citenamefont {Bose}, \citenamefont {Malik},\ and\ \citenamefont {Narain}}]{Bose_2022}%
  \BibitemOpen
  \bibfield  {author} {\bibinfo {author} {\bibfnamefont {T.}~\bibnamefont {Bose}}, \bibinfo {author} {\bibfnamefont {S.}~\bibnamefont {Malik}},\ and\ \bibinfo {author} {\bibfnamefont {M.}~\bibnamefont {Narain}},\ }\bibfield  {title} {\bibinfo {title} {{U.S. CMS - PURSUE (Program for Undergraduate Research SUmmer Experience)}},\ }in\ \href@noop {} {\emph {\bibinfo {booktitle} {{Snowmass 2021}}}}\ (\bibinfo {year} {2022})\ \Eprint {https://arxiv.org/abs/2209.10109} {arXiv:2209.10109 [physics.ed-ph]} \BibitemShut {NoStop}%
\bibitem [{\citenamefont {Banerjee}\ \emph {et~al.}(2024)\citenamefont {Banerjee}, \citenamefont {Bose}, \citenamefont {Heintz},\ and\ \citenamefont {Malik}}]{Banerjee_2024}%
  \BibitemOpen
  \bibfield  {author} {\bibinfo {author} {\bibfnamefont {S.}~\bibnamefont {Banerjee}}, \bibinfo {author} {\bibfnamefont {T.}~\bibnamefont {Bose}}, \bibinfo {author} {\bibfnamefont {U.}~\bibnamefont {Heintz}},\ and\ \bibinfo {author} {\bibfnamefont {S.}~\bibnamefont {Malik}},\ }\bibfield  {title} {\bibinfo {title} {{A novel internship program in HEP}},\ }\href@noop {} {\bibfield  {journal} {\bibinfo  {journal} {arXiv}\ } (\bibinfo {year} {2024})},\ \Eprint {https://arxiv.org/abs/2401.16217} {arXiv:2401.16217 [physics.ed-ph]} \BibitemShut {NoStop}%
\bibitem [{\citenamefont {{Murguía Burton}}\ and\ \citenamefont {Cao}(2022)}]{Murguia_2022}%
  \BibitemOpen
  \bibfield  {author} {\bibinfo {author} {\bibfnamefont {Z.~F.}\ \bibnamefont {{Murguía Burton}}}\ and\ \bibinfo {author} {\bibfnamefont {X.~E.}\ \bibnamefont {Cao}},\ }\bibfield  {title} {\bibinfo {title} {Let graduate students do internships},\ }\href {https://doi.org/https://doi.org/10.1016/j.matt.2022.11.010} {\bibfield  {journal} {\bibinfo  {journal} {Matter}\ }\textbf {\bibinfo {volume} {5}},\ \bibinfo {pages} {4100} (\bibinfo {year} {2022})}\BibitemShut {NoStop}%
\bibitem [{\citenamefont {McNeil}\ and\ \citenamefont {Heron}(2017)}]{McNeil_2017}%
  \BibitemOpen
  \bibfield  {author} {\bibinfo {author} {\bibfnamefont {L.}~\bibnamefont {McNeil}}\ and\ \bibinfo {author} {\bibfnamefont {P.}~\bibnamefont {Heron}},\ }\bibfield  {title} {\bibinfo {title} {Preparing physics students for 21st-century careers},\ }\href {https://doi.org/10.1063/PT.3.3763} {\bibfield  {journal} {\bibinfo  {journal} {Physics Today}\ }\textbf {\bibinfo {volume} {70}},\ \bibinfo {pages} {38} (\bibinfo {year} {2017})},\ \Eprint {https://arxiv.org/abs/https://pubs.aip.org/physicstoday/article-pdf/70/11/38/10118369/38\_1\_online.pdf} {https://pubs.aip.org/physicstoday/article-pdf/70/11/38/10118369/38\_1\_online.pdf} \BibitemShut {NoStop}%
\bibitem [{\citenamefont {Withheld}\ \emph {et~al.}(1995)\citenamefont {Withheld}, \citenamefont {Luhmann},\ and\ \citenamefont {Correll}}]{Withheld_1995}%
  \BibitemOpen
  \bibfield  {author} {\bibinfo {author} {\bibfnamefont {N.}~\bibnamefont {Withheld}}, \bibinfo {author} {\bibfnamefont {J.}~\bibnamefont {Luhmann}, \bibfnamefont {N.~C.}},\ and\ \bibinfo {author} {\bibfnamefont {J.}~\bibnamefont {Correll}, \bibfnamefont {D.~L.}},\ }\bibfield  {title} {\bibinfo {title} {Altering the academy in industry's interests},\ }\href {https://doi.org/10.1063/1.2808049} {\bibfield  {journal} {\bibinfo  {journal} {Physics Today}\ }\textbf {\bibinfo {volume} {48}},\ \bibinfo {pages} {13} (\bibinfo {year} {1995})},\ \Eprint {https://arxiv.org/abs/https://pubs.aip.org/physicstoday/article-pdf/48/6/13/8308512/13\_1\_online.pdf} {https://pubs.aip.org/physicstoday/article-pdf/48/6/13/8308512/13\_1\_online.pdf} \BibitemShut {NoStop}%
\bibitem [{United Nations()}]{UN2025}%
  \BibitemOpen
  United Nations,\ \href@noop {} {}\bibinfo {howpublished} {\url{https://quantum2025.org/}}\BibitemShut {NoStop}%
\bibitem [{Committee on Science and Committee on Homeland and National Security (National Science and Technology Council, Interagency Working Group on Quantum Information Science of the Subcommittee on Physical Sciences()}]{advancing_qis}%
  \BibitemOpen
  Committee on Science and Committee on Homeland and National Security (National Science and Technology Council, Interagency Working Group on Quantum Information Science of the Subcommittee on Physical Sciences,\ \href@noop {} {\bibinfo {title} {Advancing quantum information science: National challenges and opportunities}} (\bibinfo {year} {2016)})\BibitemShut {NoStop}%
\bibitem [{National Quantum Initiative Act()}]{NQIA}%
  \BibitemOpen
  National Quantum Initiative Act,\ \href {https://www.congress.gov/bill/115thcongress/house-bill/6227/text} {\bibinfo {title} {115th congress, h.r. 6227}} (\bibinfo {year} {2018})\BibitemShut {NoStop}%
\bibitem [{\citenamefont {Fox}\ \emph {et~al.}(2020)\citenamefont {Fox}, \citenamefont {Zwickl},\ and\ \citenamefont {Lewandowski}}]{Fox_2020}%
  \BibitemOpen
  \bibfield  {author} {\bibinfo {author} {\bibfnamefont {M.~F.~J.}\ \bibnamefont {Fox}}, \bibinfo {author} {\bibfnamefont {B.~M.}\ \bibnamefont {Zwickl}},\ and\ \bibinfo {author} {\bibfnamefont {H.~J.}\ \bibnamefont {Lewandowski}},\ }\bibfield  {title} {\bibinfo {title} {Preparing for the quantum revolution: What is the role of higher education?},\ }\href {https://doi.org/10.1103/PhysRevPhysEducRes.16.020131} {\bibfield  {journal} {\bibinfo  {journal} {Phys. Rev. Phys. Educ. Res.}\ }\textbf {\bibinfo {volume} {16}},\ \bibinfo {pages} {020131} (\bibinfo {year} {2020})}\BibitemShut {NoStop}%
\bibitem [{\citenamefont {{Perron}}\ \emph {et~al.}(2021)\citenamefont {{Perron}}, \citenamefont {{DeLeone}}, \citenamefont {{Sharif}}, \citenamefont {{Carter}}, \citenamefont {{Grossman}}, \citenamefont {{Passante}},\ and\ \citenamefont {{Sack}}}]{Perron_2021}%
  \BibitemOpen
  \bibfield  {author} {\bibinfo {author} {\bibfnamefont {J.~K.}\ \bibnamefont {{Perron}}}, \bibinfo {author} {\bibfnamefont {C.}~\bibnamefont {{DeLeone}}}, \bibinfo {author} {\bibfnamefont {S.}~\bibnamefont {{Sharif}}}, \bibinfo {author} {\bibfnamefont {T.}~\bibnamefont {{Carter}}}, \bibinfo {author} {\bibfnamefont {J.~M.}\ \bibnamefont {{Grossman}}}, \bibinfo {author} {\bibfnamefont {G.}~\bibnamefont {{Passante}}},\ and\ \bibinfo {author} {\bibfnamefont {J.}~\bibnamefont {{Sack}}},\ }\bibfield  {title} {\bibinfo {title} {{Quantum Undergraduate Education and Scientific Training}},\ }\href {https://doi.org/10.48550/arXiv.2109.13850} {\bibfield  {journal} {\bibinfo  {journal} {arXiv e-prints}\ ,\ \bibinfo {eid} {arXiv:2109.13850}} (\bibinfo {year} {2021})},\ \Eprint {https://arxiv.org/abs/2109.13850} {arXiv:2109.13850 [physics.ed-ph]} \BibitemShut {NoStop}%
\bibitem [{\citenamefont {Aiello}\ \emph {et~al.}(2021)\citenamefont {Aiello}, \citenamefont {Awschalom}, \citenamefont {Bernien}, \citenamefont {Brower}, \citenamefont {Brown}, \citenamefont {Brun}, \citenamefont {Caram}, \citenamefont {Chitambar}, \citenamefont {Felice}, \citenamefont {Edmonds}, \citenamefont {Fox}, \citenamefont {Haas}, \citenamefont {Holleitner}, \citenamefont {Hudson}, \citenamefont {Hunt}, \citenamefont {Joynt}, \citenamefont {Koziol}, \citenamefont {Larsen}, \citenamefont {Lewandowski}, \citenamefont {McClure}, \citenamefont {Palsberg}, \citenamefont {Passante}, \citenamefont {Pudenz}, \citenamefont {Richardson}, \citenamefont {Rosenberg}, \citenamefont {Ross}, \citenamefont {Saffman}, \citenamefont {Singh}, \citenamefont {Steuerman}, \citenamefont {Stark}, \citenamefont {Thijssen}, \citenamefont {Vamivakas}, \citenamefont {Whitfield},\ and\ \citenamefont {Zwickl}}]{Aiello_2021}%
  \BibitemOpen
  \bibfield  {author} {\bibinfo {author} {\bibfnamefont {C.~D.}\ \bibnamefont {Aiello}}, \bibinfo {author} {\bibfnamefont {D.~D.}\ \bibnamefont {Awschalom}}, \bibinfo {author} {\bibfnamefont {H.}~\bibnamefont {Bernien}}, \bibinfo {author} {\bibfnamefont {T.}~\bibnamefont {Brower}}, \bibinfo {author} {\bibfnamefont {K.~R.}\ \bibnamefont {Brown}}, \bibinfo {author} {\bibfnamefont {T.~A.}\ \bibnamefont {Brun}}, \bibinfo {author} {\bibfnamefont {J.~R.}\ \bibnamefont {Caram}}, \bibinfo {author} {\bibfnamefont {E.}~\bibnamefont {Chitambar}}, \bibinfo {author} {\bibfnamefont {R.~D.}\ \bibnamefont {Felice}}, \bibinfo {author} {\bibfnamefont {K.~M.}\ \bibnamefont {Edmonds}}, \bibinfo {author} {\bibfnamefont {M.~F.~J.}\ \bibnamefont {Fox}}, \bibinfo {author} {\bibfnamefont {S.}~\bibnamefont {Haas}}, \bibinfo {author} {\bibfnamefont {A.~W.}\ \bibnamefont {Holleitner}}, \bibinfo {author} {\bibfnamefont {E.~R.}\ \bibnamefont {Hudson}}, \bibinfo {author} {\bibfnamefont {J.~H.}\ \bibnamefont {Hunt}}, \bibinfo {author}
  {\bibfnamefont {R.}~\bibnamefont {Joynt}}, \bibinfo {author} {\bibfnamefont {S.}~\bibnamefont {Koziol}}, \bibinfo {author} {\bibfnamefont {M.}~\bibnamefont {Larsen}}, \bibinfo {author} {\bibfnamefont {H.~J.}\ \bibnamefont {Lewandowski}}, \bibinfo {author} {\bibfnamefont {D.~T.}\ \bibnamefont {McClure}}, \bibinfo {author} {\bibfnamefont {J.}~\bibnamefont {Palsberg}}, \bibinfo {author} {\bibfnamefont {G.}~\bibnamefont {Passante}}, \bibinfo {author} {\bibfnamefont {K.~L.}\ \bibnamefont {Pudenz}}, \bibinfo {author} {\bibfnamefont {C.~J.~K.}\ \bibnamefont {Richardson}}, \bibinfo {author} {\bibfnamefont {J.~L.}\ \bibnamefont {Rosenberg}}, \bibinfo {author} {\bibfnamefont {R.~S.}\ \bibnamefont {Ross}}, \bibinfo {author} {\bibfnamefont {M.}~\bibnamefont {Saffman}}, \bibinfo {author} {\bibfnamefont {M.}~\bibnamefont {Singh}}, \bibinfo {author} {\bibfnamefont {D.~W.}\ \bibnamefont {Steuerman}}, \bibinfo {author} {\bibfnamefont {C.}~\bibnamefont {Stark}}, \bibinfo {author} {\bibfnamefont {J.}~\bibnamefont {Thijssen}},
  \bibinfo {author} {\bibfnamefont {A.~N.}\ \bibnamefont {Vamivakas}}, \bibinfo {author} {\bibfnamefont {J.~D.}\ \bibnamefont {Whitfield}},\ and\ \bibinfo {author} {\bibfnamefont {B.~M.}\ \bibnamefont {Zwickl}},\ }\bibfield  {title} {\bibinfo {title} {Achieving a quantum smart workforce},\ }\href {https://doi.org/10.1088/2058-9565/abfa64} {\bibfield  {journal} {\bibinfo  {journal} {Quantum Science and Technology}\ }\textbf {\bibinfo {volume} {6}},\ \bibinfo {pages} {030501} (\bibinfo {year} {2021})}\BibitemShut {NoStop}%
\bibitem [{\citenamefont {Hughes}\ \emph {et~al.}(2022)\citenamefont {Hughes}, \citenamefont {Finke}, \citenamefont {German}, \citenamefont {Merzbacher}, \citenamefont {Vora},\ and\ \citenamefont {Lewandowski}}]{Hughes_2022}%
  \BibitemOpen
  \bibfield  {author} {\bibinfo {author} {\bibfnamefont {C.}~\bibnamefont {Hughes}}, \bibinfo {author} {\bibfnamefont {D.}~\bibnamefont {Finke}}, \bibinfo {author} {\bibfnamefont {D.-A.}\ \bibnamefont {German}}, \bibinfo {author} {\bibfnamefont {C.}~\bibnamefont {Merzbacher}}, \bibinfo {author} {\bibfnamefont {P.~M.}\ \bibnamefont {Vora}},\ and\ \bibinfo {author} {\bibfnamefont {H.~J.}\ \bibnamefont {Lewandowski}},\ }\bibfield  {title} {\bibinfo {title} {Assessing the needs of the quantum industry},\ }\href {https://doi.org/10.1109/TE.2022.3153841} {\bibfield  {journal} {\bibinfo  {journal} {IEEE Transactions on Education}\ }\textbf {\bibinfo {volume} {65}},\ \bibinfo {pages} {592} (\bibinfo {year} {2022})}\BibitemShut {NoStop}%
\bibitem [{\citenamefont {{Asfaw}}\ \emph {et~al.}(2022)\citenamefont {{Asfaw}}, \citenamefont {{Blais}}, \citenamefont {{Brown}}, \citenamefont {{Candelaria}}, \citenamefont {{Cantwell}}, \citenamefont {{Carr}}, \citenamefont {{Combes}}, \citenamefont {{Debroy}}, \citenamefont {{Donohue}}, \citenamefont {{Economou}}, \citenamefont {{Edwards}}, \citenamefont {{Fox}}, \citenamefont {{Girvin}}, \citenamefont {{Ho}}, \citenamefont {{Hurst}}, \citenamefont {{Jacob}}, \citenamefont {{Johnson}}, \citenamefont {{Johnston-Halperin}}, \citenamefont {{Joynt}}, \citenamefont {{Kapit}}, \citenamefont {{Klein-Seetharaman}}, \citenamefont {{Laforest}}, \citenamefont {{Lewandowski}}, \citenamefont {{Lynn}}, \citenamefont {{McRae}}, \citenamefont {{Merzbacher}}, \citenamefont {{Michalakis}}, \citenamefont {{Narang}}, \citenamefont {{Oliver}}, \citenamefont {{Palsberg}}, \citenamefont {{Pappas}}, \citenamefont {{Raymer}}, \citenamefont {{Reilly}}, \citenamefont {{Saffman}}, \citenamefont {{Searles}}, \citenamefont
  {{Shapiro}},\ and\ \citenamefont {{Singh}}}]{Asfaw_2022}%
  \BibitemOpen
  \bibfield  {author} {\bibinfo {author} {\bibfnamefont {A.}~\bibnamefont {{Asfaw}}}, \bibinfo {author} {\bibfnamefont {A.}~\bibnamefont {{Blais}}}, \bibinfo {author} {\bibfnamefont {K.~R.}\ \bibnamefont {{Brown}}}, \bibinfo {author} {\bibfnamefont {J.}~\bibnamefont {{Candelaria}}}, \bibinfo {author} {\bibfnamefont {C.}~\bibnamefont {{Cantwell}}}, \bibinfo {author} {\bibfnamefont {L.~D.}\ \bibnamefont {{Carr}}}, \bibinfo {author} {\bibfnamefont {J.}~\bibnamefont {{Combes}}}, \bibinfo {author} {\bibfnamefont {D.~M.}\ \bibnamefont {{Debroy}}}, \bibinfo {author} {\bibfnamefont {J.~M.}\ \bibnamefont {{Donohue}}}, \bibinfo {author} {\bibfnamefont {S.~E.}\ \bibnamefont {{Economou}}}, \bibinfo {author} {\bibfnamefont {E.}~\bibnamefont {{Edwards}}}, \bibinfo {author} {\bibfnamefont {M.~F.~J.}\ \bibnamefont {{Fox}}}, \bibinfo {author} {\bibfnamefont {S.~M.}\ \bibnamefont {{Girvin}}}, \bibinfo {author} {\bibfnamefont {A.}~\bibnamefont {{Ho}}}, \bibinfo {author} {\bibfnamefont {H.~M.}\ \bibnamefont {{Hurst}}}, \bibinfo
  {author} {\bibfnamefont {Z.}~\bibnamefont {{Jacob}}}, \bibinfo {author} {\bibfnamefont {B.~R.}\ \bibnamefont {{Johnson}}}, \bibinfo {author} {\bibfnamefont {E.}~\bibnamefont {{Johnston-Halperin}}}, \bibinfo {author} {\bibfnamefont {R.}~\bibnamefont {{Joynt}}}, \bibinfo {author} {\bibfnamefont {E.}~\bibnamefont {{Kapit}}}, \bibinfo {author} {\bibfnamefont {J.}~\bibnamefont {{Klein-Seetharaman}}}, \bibinfo {author} {\bibfnamefont {M.}~\bibnamefont {{Laforest}}}, \bibinfo {author} {\bibfnamefont {H.~J.}\ \bibnamefont {{Lewandowski}}}, \bibinfo {author} {\bibfnamefont {T.~W.}\ \bibnamefont {{Lynn}}}, \bibinfo {author} {\bibfnamefont {C.~R.~H.}\ \bibnamefont {{McRae}}}, \bibinfo {author} {\bibfnamefont {C.}~\bibnamefont {{Merzbacher}}}, \bibinfo {author} {\bibfnamefont {S.}~\bibnamefont {{Michalakis}}}, \bibinfo {author} {\bibfnamefont {P.}~\bibnamefont {{Narang}}}, \bibinfo {author} {\bibfnamefont {W.~D.}\ \bibnamefont {{Oliver}}}, \bibinfo {author} {\bibfnamefont {J.}~\bibnamefont {{Palsberg}}}, \bibinfo
  {author} {\bibfnamefont {D.~P.}\ \bibnamefont {{Pappas}}}, \bibinfo {author} {\bibfnamefont {M.~G.}\ \bibnamefont {{Raymer}}}, \bibinfo {author} {\bibfnamefont {D.~J.}\ \bibnamefont {{Reilly}}}, \bibinfo {author} {\bibfnamefont {M.}~\bibnamefont {{Saffman}}}, \bibinfo {author} {\bibfnamefont {T.~A.}\ \bibnamefont {{Searles}}}, \bibinfo {author} {\bibfnamefont {J.~H.}\ \bibnamefont {{Shapiro}}},\ and\ \bibinfo {author} {\bibfnamefont {C.}~\bibnamefont {{Singh}}},\ }\bibfield  {title} {\bibinfo {title} {{Building a Quantum Engineering Undergraduate Program}},\ }\href {https://doi.org/10.1109/TE.2022.3144943} {\bibfield  {journal} {\bibinfo  {journal} {IEEE Transactions on Education}\ }\textbf {\bibinfo {volume} {65}},\ \bibinfo {pages} {220} (\bibinfo {year} {2022})},\ \Eprint {https://arxiv.org/abs/2108.01311} {arXiv:2108.01311 [physics.ed-ph]} \BibitemShut {NoStop}%
\bibitem [{\citenamefont {Kaur}\ and\ \citenamefont {Venegas-Gomez}(2022)}]{Maninder_2022}%
  \BibitemOpen
  \bibfield  {author} {\bibinfo {author} {\bibfnamefont {M.}~\bibnamefont {Kaur}}\ and\ \bibinfo {author} {\bibfnamefont {A.}~\bibnamefont {Venegas-Gomez}},\ }\bibfield  {title} {\bibinfo {title} {{Defining the quantum workforce landscape: a review of global quantum education initiatives}},\ }\href {https://doi.org/10.1117/1.OE.61.8.081806} {\bibfield  {journal} {\bibinfo  {journal} {Optical Engineering}\ }\textbf {\bibinfo {volume} {61}},\ \bibinfo {pages} {081806} (\bibinfo {year} {2022})}\BibitemShut {NoStop}%
\bibitem [{\citenamefont {Raghunathan}\ \emph {et~al.}(2023)\citenamefont {Raghunathan}, \citenamefont {Chekuri}, \citenamefont {Fuhrer}, \citenamefont {Wildemann}, \citenamefont {Wu}, \citenamefont {Sarabi}, \citenamefont {Scales}, \citenamefont {Lester},\ and\ \citenamefont {Kovanis}}]{Raghunathan_2023}%
  \BibitemOpen
  \bibfield  {author} {\bibinfo {author} {\bibfnamefont {R.}~\bibnamefont {Raghunathan}}, \bibinfo {author} {\bibfnamefont {V.~A.}\ \bibnamefont {Chekuri}}, \bibinfo {author} {\bibfnamefont {M.}~\bibnamefont {Fuhrer}}, \bibinfo {author} {\bibfnamefont {I.}~\bibnamefont {Wildemann}}, \bibinfo {author} {\bibfnamefont {C.}~\bibnamefont {Wu}}, \bibinfo {author} {\bibfnamefont {B.}~\bibnamefont {Sarabi}}, \bibinfo {author} {\bibfnamefont {W.}~\bibnamefont {Scales}}, \bibinfo {author} {\bibfnamefont {L.~F.}\ \bibnamefont {Lester}},\ and\ \bibinfo {author} {\bibfnamefont {V.}~\bibnamefont {Kovanis}},\ }\bibfield  {title} {\bibinfo {title} {{Toward a quantum-enabled workforce: curriculum design, project based learning, and institutional partnerships as integrated methodologies}},\ }in\ \href {https://doi.org/10.1117/12.2673044} {\emph {\bibinfo {booktitle} {Seventeenth Conference on Education and Training in Optics and Photonics: ETOP 2023}}},\ Vol.\ \bibinfo {volume} {12723},\ \bibinfo {editor} {edited by\ \bibinfo
  {editor} {\bibfnamefont {M.}~\bibnamefont {McKee}}\ and\ \bibinfo {editor} {\bibfnamefont {D.~J.}\ \bibnamefont {Hagan}}},\ \bibinfo {organization} {International Society for Optics and Photonics}\ (\bibinfo  {publisher} {SPIE},\ \bibinfo {year} {2023})\ p.\ \bibinfo {pages} {1272335}\BibitemShut {NoStop}%
\bibitem [{\citenamefont {Darienzo}\ \emph {et~al.}(2024)\citenamefont {Darienzo}, \citenamefont {Kelly}, \citenamefont {Schneble},\ and\ \citenamefont {Wei}}]{Darienzo_2024}%
  \BibitemOpen
  \bibfield  {author} {\bibinfo {author} {\bibfnamefont {M.}~\bibnamefont {Darienzo}}, \bibinfo {author} {\bibfnamefont {A.~M.}\ \bibnamefont {Kelly}}, \bibinfo {author} {\bibfnamefont {D.}~\bibnamefont {Schneble}},\ and\ \bibinfo {author} {\bibfnamefont {T.-C.}\ \bibnamefont {Wei}},\ }\bibfield  {title} {\bibinfo {title} {Student attitudes toward quantum information science and technology in a high school outreach program},\ }\href {https://doi.org/10.1103/PhysRevPhysEducRes.20.020126} {\bibfield  {journal} {\bibinfo  {journal} {Phys. Rev. Phys. Educ. Res.}\ }\textbf {\bibinfo {volume} {20}},\ \bibinfo {pages} {020126} (\bibinfo {year} {2024})}\BibitemShut {NoStop}%
\bibitem [{QLCI - Quantum Leap Challenge Institutes()}]{qlci}%
  \BibitemOpen
  QLCI - Quantum Leap Challenge Institutes,\ \href@noop {} {}\bibinfo {howpublished} {https://www.nsf.gov/funding/opportunities/qlci-quantum-leap-challenge-institutes}\BibitemShut {NoStop}%
\bibitem [{Q-Sense - Quantum Sensing through Entangled Science and Engineering.()}]{qsense}%
  \BibitemOpen
  Q-Sense - Quantum Sensing through Entangled Science and Engineering.,\ \href@noop {} {}\bibinfo {howpublished} {https://www.colorado.edu/research/qsense/}\BibitemShut {NoStop}%
\bibitem [{\citenamefont {{Bennett}}\ \emph {et~al.}(2024)\citenamefont {{Bennett}}, \citenamefont {{Arrow}}, \citenamefont {{Novack}},\ and\ \citenamefont {{Finkelstein}}}]{Bennett_2024}%
  \BibitemOpen
  \bibfield  {author} {\bibinfo {author} {\bibfnamefont {M.~B.}\ \bibnamefont {{Bennett}}}, \bibinfo {author} {\bibfnamefont {J.~{\'E}.}\ \bibnamefont {{Arrow}}}, \bibinfo {author} {\bibfnamefont {S.}~\bibnamefont {{Novack}}},\ and\ \bibinfo {author} {\bibfnamefont {N.~D.}\ \bibnamefont {{Finkelstein}}},\ }\bibfield  {title} {\bibinfo {title} {{Investigating Student Participation in Quantum Workforce Initiatives}},\ }\href {https://doi.org/10.48550/arXiv.2407.14698} {\bibfield  {journal} {\bibinfo  {journal} {arXiv e-prints}\ ,\ \bibinfo {eid} {arXiv:2407.14698}} (\bibinfo {year} {2024})},\ \Eprint {https://arxiv.org/abs/2407.14698} {arXiv:2407.14698 [physics.ed-ph]} \BibitemShut {NoStop}%
\bibitem [{EdQuantum - Hybrid Curriculum in Advenced Optics, Spectroscopy, and Quantum Technologies.()}]{edquantum}%
  \BibitemOpen
  EdQuantum - Hybrid Curriculum in Advenced Optics, Spectroscopy, and Quantum Technologies.,\ \href@noop {} {}\bibinfo {howpublished} {https://edquantum.org/}\BibitemShut {NoStop}%
\bibitem [{\citenamefont {Hasanovic}\ \emph {et~al.}(2022)\citenamefont {Hasanovic}, \citenamefont {Panayiotou}, \citenamefont {Silberman}, \citenamefont {Stimers},\ and\ \citenamefont {Merzbacher}}]{Hasanovic_2022}%
  \BibitemOpen
  \bibfield  {author} {\bibinfo {author} {\bibfnamefont {M.}~\bibnamefont {Hasanovic}}, \bibinfo {author} {\bibfnamefont {C.~A.}\ \bibnamefont {Panayiotou}}, \bibinfo {author} {\bibfnamefont {D.~M.}\ \bibnamefont {Silberman}}, \bibinfo {author} {\bibfnamefont {P.}~\bibnamefont {Stimers}},\ and\ \bibinfo {author} {\bibfnamefont {C.~I.}\ \bibnamefont {Merzbacher}},\ }\bibfield  {title} {\bibinfo {title} {{Quantum technician skills and competencies for the emerging Quantum 2.0 industry}},\ }\href {https://doi.org/10.1117/1.OE.61.8.081803} {\bibfield  {journal} {\bibinfo  {journal} {Optical Engineering}\ }\textbf {\bibinfo {volume} {61}},\ \bibinfo {pages} {081803} (\bibinfo {year} {2022})}\BibitemShut {NoStop}%
\bibitem [{\citenamefont {{Oliver}}\ \emph {et~al.}(2024)\citenamefont {{Oliver}}, \citenamefont {{Borish}}, \citenamefont {{Wilcox}},\ and\ \citenamefont {{Lewandowski}}}]{Oliver_2024}%
  \BibitemOpen
  \bibfield  {author} {\bibinfo {author} {\bibfnamefont {K.~A.}\ \bibnamefont {{Oliver}}}, \bibinfo {author} {\bibfnamefont {V.}~\bibnamefont {{Borish}}}, \bibinfo {author} {\bibfnamefont {B.~R.}\ \bibnamefont {{Wilcox}}},\ and\ \bibinfo {author} {\bibfnamefont {H.~J.}\ \bibnamefont {{Lewandowski}}},\ }\bibfield  {title} {\bibinfo {title} {{Education for expanding the quantum workforce: Student perceptions of the quantum industry in an upper-division physics capstone course}},\ }\href {https://doi.org/10.48550/arXiv.2407.07902} {\bibfield  {journal} {\bibinfo  {journal} {arXiv e-prints}\ ,\ \bibinfo {eid} {arXiv:2407.07902}} (\bibinfo {year} {2024})},\ \Eprint {https://arxiv.org/abs/2407.07902} {arXiv:2407.07902 [physics.ed-ph]} \BibitemShut {NoStop}%
\bibitem [{Quantum Technology Education.()}]{qtedu}%
  \BibitemOpen
  Quantum Technology Education.,\ \href@noop {} {}\bibinfo {howpublished} {https://qtedu.eu/}\BibitemShut {NoStop}%
\bibitem [{\citenamefont {Gerke}\ \emph {et~al.}(2022)\citenamefont {Gerke}, \citenamefont {Müller}, \citenamefont {Bitzenbauer}, \citenamefont {Ubben},\ and\ \citenamefont {Weber}}]{Gerke_2022}%
  \BibitemOpen
  \bibfield  {author} {\bibinfo {author} {\bibfnamefont {F.}~\bibnamefont {Gerke}}, \bibinfo {author} {\bibfnamefont {R.}~\bibnamefont {Müller}}, \bibinfo {author} {\bibfnamefont {P.}~\bibnamefont {Bitzenbauer}}, \bibinfo {author} {\bibfnamefont {M.}~\bibnamefont {Ubben}},\ and\ \bibinfo {author} {\bibfnamefont {K.-A.}\ \bibnamefont {Weber}},\ }\bibfield  {title} {\bibinfo {title} {Requirements for future quantum workforce – a delphi study},\ }\href {https://doi.org/10.1088/1742-6596/2297/1/012017} {\bibfield  {journal} {\bibinfo  {journal} {Journal of Physics: Conference Series}\ }\textbf {\bibinfo {volume} {2297}},\ \bibinfo {pages} {012017} (\bibinfo {year} {2022})}\BibitemShut {NoStop}%
\bibitem [{QUCATS - Coordination and support action of the Quantum Flagship.()}]{qucats}%
  \BibitemOpen
  QUCATS - Coordination and support action of the Quantum Flagship.,\ \href@noop {} {}\bibinfo {howpublished} {https://qt.eu/projects/csa-projects/qucats}\BibitemShut {NoStop}%
\bibitem [{\citenamefont {Greinert}\ \emph {et~al.}(2023)\citenamefont {Greinert}, \citenamefont {M\"uller}, \citenamefont {Bitzenbauer}, \citenamefont {Ubben},\ and\ \citenamefont {Weber}}]{Greinert_2023}%
  \BibitemOpen
  \bibfield  {author} {\bibinfo {author} {\bibfnamefont {F.}~\bibnamefont {Greinert}}, \bibinfo {author} {\bibfnamefont {R.}~\bibnamefont {M\"uller}}, \bibinfo {author} {\bibfnamefont {P.}~\bibnamefont {Bitzenbauer}}, \bibinfo {author} {\bibfnamefont {M.~S.}\ \bibnamefont {Ubben}},\ and\ \bibinfo {author} {\bibfnamefont {K.-A.}\ \bibnamefont {Weber}},\ }\bibfield  {title} {\bibinfo {title} {Future quantum workforce: Competences, requirements, and forecasts},\ }\href {https://doi.org/10.1103/PhysRevPhysEducRes.19.010137} {\bibfield  {journal} {\bibinfo  {journal} {Phys. Rev. Phys. Educ. Res.}\ }\textbf {\bibinfo {volume} {19}},\ \bibinfo {pages} {010137} (\bibinfo {year} {2023})}\BibitemShut {NoStop}%
\bibitem [{\citenamefont {F}\ \emph {et~al.}(2023)\citenamefont {F}, \citenamefont {R}, \citenamefont {S}, \citenamefont {J},\ and\ \citenamefont {MS}}]{Greinert_2023_2}%
  \BibitemOpen
  \bibfield  {author} {\bibinfo {author} {\bibfnamefont {G.}~\bibnamefont {F}}, \bibinfo {author} {\bibfnamefont {M.}~\bibnamefont {R}}, \bibinfo {author} {\bibfnamefont {G.}~\bibnamefont {S}}, \bibinfo {author} {\bibfnamefont {S.}~\bibnamefont {J}},\ and\ \bibinfo {author} {\bibfnamefont {U.}~\bibnamefont {MS}},\ }\bibfield  {title} {\bibinfo {title} {Towards a quantum ready workforce: the updated european competence framework for quantum technologies},\ }\bibfield  {journal} {\bibinfo  {journal} {Front. Quantum Sci. Technol.}\ }\href {https://doi.org/10.3389/frqst.2023.1225733} {10.3389/frqst.2023.1225733} (\bibinfo {year} {2023})\BibitemShut {NoStop}%
\bibitem [{DigiQ - Digitally Enhanced Quantum Technology Master.()}]{digiq}%
  \BibitemOpen
  DigiQ - Digitally Enhanced Quantum Technology Master.,\ \href@noop {} {}\bibinfo {howpublished} {https://www.digiq.eu/}\BibitemShut {NoStop}%
\bibitem [{Quantum Technologies Courses for Industry.()}]{qtindu}%
  \BibitemOpen
  Quantum Technologies Courses for Industry.,\ \href@noop {} {}\bibinfo {howpublished} {https://qtindu.eu/}\BibitemShut {NoStop}%
\bibitem [{\citenamefont {Greinert}\ \emph {et~al.}(2024)\citenamefont {Greinert}, \citenamefont {Ubben}, \citenamefont {Dogan}, \citenamefont {Hilfert-Rüppell},\ and\ \citenamefont {Müller}}]{Greinert_2024}%
  \BibitemOpen
  \bibfield  {author} {\bibinfo {author} {\bibfnamefont {F.}~\bibnamefont {Greinert}}, \bibinfo {author} {\bibfnamefont {M.~S.}\ \bibnamefont {Ubben}}, \bibinfo {author} {\bibfnamefont {I.~N.}\ \bibnamefont {Dogan}}, \bibinfo {author} {\bibfnamefont {D.}~\bibnamefont {Hilfert-Rüppell}},\ and\ \bibinfo {author} {\bibfnamefont {R.}~\bibnamefont {Müller}},\ }\bibfield  {title} {\bibinfo {title} {Advancing quantum technology workforce: industry insights into qualification and training needs},\ }\bibfield  {journal} {\bibinfo  {journal} {EPJ Quantum Technology}\ }\textbf {\bibinfo {volume} {11}},\ \href {https://doi.org/10.1140/epjqt/s40507-024-00294-2} {10.1140/epjqt/s40507-024-00294-2} (\bibinfo {year} {2024})\BibitemShut {NoStop}%
\bibitem [{\citenamefont {Biswas}\ \emph {et~al.}(2017)\citenamefont {Biswas}, \citenamefont {Jiang}, \citenamefont {Kechezhi}, \citenamefont {Knysh}, \citenamefont {Mandrà}, \citenamefont {O’Gorman}, \citenamefont {Perdomo-Ortiz}, \citenamefont {Petukhov}, \citenamefont {Realpe-Gómez}, \citenamefont {Rieffel}, \citenamefont {Venturelli}, \citenamefont {Vasko},\ and\ \citenamefont {Wang}}]{Biswas_2017}%
  \BibitemOpen
  \bibfield  {author} {\bibinfo {author} {\bibfnamefont {R.}~\bibnamefont {Biswas}}, \bibinfo {author} {\bibfnamefont {Z.}~\bibnamefont {Jiang}}, \bibinfo {author} {\bibfnamefont {K.}~\bibnamefont {Kechezhi}}, \bibinfo {author} {\bibfnamefont {S.}~\bibnamefont {Knysh}}, \bibinfo {author} {\bibfnamefont {S.}~\bibnamefont {Mandrà}}, \bibinfo {author} {\bibfnamefont {B.}~\bibnamefont {O’Gorman}}, \bibinfo {author} {\bibfnamefont {A.}~\bibnamefont {Perdomo-Ortiz}}, \bibinfo {author} {\bibfnamefont {A.}~\bibnamefont {Petukhov}}, \bibinfo {author} {\bibfnamefont {J.}~\bibnamefont {Realpe-Gómez}}, \bibinfo {author} {\bibfnamefont {E.}~\bibnamefont {Rieffel}}, \bibinfo {author} {\bibfnamefont {D.}~\bibnamefont {Venturelli}}, \bibinfo {author} {\bibfnamefont {F.}~\bibnamefont {Vasko}},\ and\ \bibinfo {author} {\bibfnamefont {Z.}~\bibnamefont {Wang}},\ }\bibfield  {title} {\bibinfo {title} {A nasa perspective on quantum computing: Opportunities and challenges},\ }\href
  {https://doi.org/https://doi.org/10.1016/j.parco.2016.11.002} {\bibfield  {journal} {\bibinfo  {journal} {Parallel Computing}\ }\textbf {\bibinfo {volume} {64}},\ \bibinfo {pages} {81} (\bibinfo {year} {2017})},\ \bibinfo {note} {high-End Computing for Next-Generation Scientific Discovery}\BibitemShut {NoStop}%
\bibitem [{Los Alamos National Laboratory()}]{los_alamos}%
  \BibitemOpen
  Los Alamos National Laboratory,\ \href@noop {} {}\bibinfo {howpublished} {\url{https://www.lanl.gov/engage/collaboration/internships/summer-schools/quantumschool}}\BibitemShut {NoStop}%
\bibitem [{Fermilab()}]{sqms}%
  \BibitemOpen
  Fermilab,\ \href@noop {} {}\bibinfo {howpublished} {\url{https://sqmscenter.fnal.gov/opportunities/internships-and-fellowships/}}\BibitemShut {NoStop}%
\bibitem [{\citenamefont {Alam}\ \emph {et~al.}(2022)\citenamefont {Alam}, \citenamefont {Hadfield}, \citenamefont {Lamm},\ and\ \citenamefont {Li}}]{Alam_2022}%
  \BibitemOpen
  \bibfield  {author} {\bibinfo {author} {\bibfnamefont {M.~S.}\ \bibnamefont {Alam}}, \bibinfo {author} {\bibfnamefont {S.}~\bibnamefont {Hadfield}}, \bibinfo {author} {\bibfnamefont {H.}~\bibnamefont {Lamm}},\ and\ \bibinfo {author} {\bibfnamefont {A.~C.}\ \bibnamefont {Li}},\ }\bibfield  {title} {\bibinfo {title} {Primitive quantum gates for dihedral gauge theories},\ }\bibfield  {journal} {\bibinfo  {journal} {Physical Review D}\ }\textbf {\bibinfo {volume} {105}},\ \href {https://doi.org/10.1103/physrevd.105.114501} {10.1103/physrevd.105.114501} (\bibinfo {year} {2022})\BibitemShut {NoStop}%
\bibitem [{\citenamefont {Bassman}\ \emph {et~al.}(2021)\citenamefont {Bassman}, \citenamefont {Klymko}, \citenamefont {Liu}, \citenamefont {Tubman},\ and\ \citenamefont {de~Jong}}]{bassman2021computing}%
  \BibitemOpen
  \bibfield  {author} {\bibinfo {author} {\bibfnamefont {L.}~\bibnamefont {Bassman}}, \bibinfo {author} {\bibfnamefont {K.}~\bibnamefont {Klymko}}, \bibinfo {author} {\bibfnamefont {D.}~\bibnamefont {Liu}}, \bibinfo {author} {\bibfnamefont {N.~M.}\ \bibnamefont {Tubman}},\ and\ \bibinfo {author} {\bibfnamefont {W.~A.}\ \bibnamefont {de~Jong}},\ }\href@noop {} {\bibinfo {title} {Computing free energies with fluctuation relations on quantum computers}} (\bibinfo {year} {2021}),\ \Eprint {https://arxiv.org/abs/2103.09846} {arXiv:2103.09846 [quant-ph]} \BibitemShut {NoStop}%
\bibitem [{\citenamefont {Kerger}\ and\ \citenamefont {Miyazaki}(2023)}]{Kerger2023}%
  \BibitemOpen
  \bibfield  {author} {\bibinfo {author} {\bibfnamefont {P.}~\bibnamefont {Kerger}}\ and\ \bibinfo {author} {\bibfnamefont {R.}~\bibnamefont {Miyazaki}},\ }\bibfield  {title} {\bibinfo {title} {Quantum image denoising: a framework via boltzmann machines, qubo, and quantum annealing},\ }\bibfield  {journal} {\bibinfo  {journal} {Frontiers in Computer Science}\ }\textbf {\bibinfo {volume} {5}},\ \href {https://doi.org/10.3389/fcomp.2023.1281100} {10.3389/fcomp.2023.1281100} (\bibinfo {year} {2023})\BibitemShut {NoStop}%
\bibitem [{\citenamefont {Juenger}\ \emph {et~al.}(2019)\citenamefont {Juenger}, \citenamefont {Lobe}, \citenamefont {Mutzel}, \citenamefont {Reinelt}, \citenamefont {Rendl}, \citenamefont {Rinaldi},\ and\ \citenamefont {Stollenwerk}}]{juenger2019performance}%
  \BibitemOpen
  \bibfield  {author} {\bibinfo {author} {\bibfnamefont {M.}~\bibnamefont {Juenger}}, \bibinfo {author} {\bibfnamefont {E.}~\bibnamefont {Lobe}}, \bibinfo {author} {\bibfnamefont {P.}~\bibnamefont {Mutzel}}, \bibinfo {author} {\bibfnamefont {G.}~\bibnamefont {Reinelt}}, \bibinfo {author} {\bibfnamefont {F.}~\bibnamefont {Rendl}}, \bibinfo {author} {\bibfnamefont {G.}~\bibnamefont {Rinaldi}},\ and\ \bibinfo {author} {\bibfnamefont {T.}~\bibnamefont {Stollenwerk}},\ }\href@noop {} {\bibinfo {title} {Performance of a quantum annealer for ising ground state computations on chimera graphs}} (\bibinfo {year} {2019}),\ \Eprint {https://arxiv.org/abs/1904.11965} {arXiv:1904.11965 [cs.DS]} \BibitemShut {NoStop}%
\bibitem [{\citenamefont {Gonzalez~Izquierdo}\ \emph {et~al.}(2021)\citenamefont {Gonzalez~Izquierdo}, \citenamefont {Grabbe}, \citenamefont {Hadfield}, \citenamefont {Marshall}, \citenamefont {Wang},\ and\ \citenamefont {Rieffel}}]{Gonzalez_Izquierdo_2021}%
  \BibitemOpen
  \bibfield  {author} {\bibinfo {author} {\bibfnamefont {Z.}~\bibnamefont {Gonzalez~Izquierdo}}, \bibinfo {author} {\bibfnamefont {S.}~\bibnamefont {Grabbe}}, \bibinfo {author} {\bibfnamefont {S.}~\bibnamefont {Hadfield}}, \bibinfo {author} {\bibfnamefont {J.}~\bibnamefont {Marshall}}, \bibinfo {author} {\bibfnamefont {Z.}~\bibnamefont {Wang}},\ and\ \bibinfo {author} {\bibfnamefont {E.}~\bibnamefont {Rieffel}},\ }\bibfield  {title} {\bibinfo {title} {Ferromagnetically shifting the power of pausing},\ }\bibfield  {journal} {\bibinfo  {journal} {Physical Review Applied}\ }\textbf {\bibinfo {volume} {15}},\ \href {https://doi.org/10.1103/physrevapplied.15.044013} {10.1103/physrevapplied.15.044013} (\bibinfo {year} {2021})\BibitemShut {NoStop}%
\bibitem [{\citenamefont {Izquierdo}\ \emph {et~al.}(2021)\citenamefont {Izquierdo}, \citenamefont {Hen},\ and\ \citenamefont {Albash}}]{Izquierdo_2021}%
  \BibitemOpen
  \bibfield  {author} {\bibinfo {author} {\bibfnamefont {Z.~G.}\ \bibnamefont {Izquierdo}}, \bibinfo {author} {\bibfnamefont {I.}~\bibnamefont {Hen}},\ and\ \bibinfo {author} {\bibfnamefont {T.}~\bibnamefont {Albash}},\ }\bibfield  {title} {\bibinfo {title} {Testing a quantum annealer as a quantum thermal sampler},\ }\href {https://doi.org/10.1145/3464456} {\bibfield  {journal} {\bibinfo  {journal} {ACM Transactions on Quantum Computing}\ }\textbf {\bibinfo {volume} {2}},\ \bibinfo {pages} {1–20} (\bibinfo {year} {2021})}\BibitemShut {NoStop}%
\bibitem [{\citenamefont {Pokharel}\ \emph {et~al.}(2023)\citenamefont {Pokharel}, \citenamefont {Gonzalez~Izquierdo},\ and\ \citenamefont {Lott}}]{Pokharel_2023}%
  \BibitemOpen
  \bibfield  {author} {\bibinfo {author} {\bibfnamefont {B.}~\bibnamefont {Pokharel}}, \bibinfo {author} {\bibfnamefont {Z.}~\bibnamefont {Gonzalez~Izquierdo}},\ and\ \bibinfo {author} {\bibfnamefont {P.~e.~a.}\ \bibnamefont {Lott}},\ }\bibfield  {title} {\bibinfo {title} {Inter-generational comparison of quantum annealers in solving hard scheduling problems},\ }\href {https://doi.org/10.1007/s11128-023-04077-z} {\bibfield  {journal} {\bibinfo  {journal} {Quantum Information Processing}\ }\textbf {\bibinfo {volume} {22}} (\bibinfo {year} {2023})}\BibitemShut {NoStop}%
\bibitem [{\citenamefont {Bernal}\ \emph {et~al.}(2019)\citenamefont {Bernal}, \citenamefont {Booth}, \citenamefont {Dridi}, \citenamefont {Alghassi}, \citenamefont {Tayur},\ and\ \citenamefont {Venturelli}}]{bernal2019integer}%
  \BibitemOpen
  \bibfield  {author} {\bibinfo {author} {\bibfnamefont {D.~E.}\ \bibnamefont {Bernal}}, \bibinfo {author} {\bibfnamefont {K.~E.~C.}\ \bibnamefont {Booth}}, \bibinfo {author} {\bibfnamefont {R.}~\bibnamefont {Dridi}}, \bibinfo {author} {\bibfnamefont {H.}~\bibnamefont {Alghassi}}, \bibinfo {author} {\bibfnamefont {S.}~\bibnamefont {Tayur}},\ and\ \bibinfo {author} {\bibfnamefont {D.}~\bibnamefont {Venturelli}},\ }\href@noop {} {\bibinfo {title} {Integer programming techniques for minor-embedding in quantum annealers}} (\bibinfo {year} {2019}),\ \Eprint {https://arxiv.org/abs/1912.08314} {arXiv:1912.08314 [quant-ph]} \BibitemShut {NoStop}%
\bibitem [{\citenamefont {{Levy}}\ \emph {et~al.}(2022)\citenamefont {{Levy}}, \citenamefont {{Gonzalez Izquierdo}}, \citenamefont {{Wang}}, \citenamefont {{Marshall}}, \citenamefont {{Barreto}}, \citenamefont {{Fry-Bouriaux}}, \citenamefont {{O'Connor}}, \citenamefont {{Warburton}}, \citenamefont {{Wiebe}}, \citenamefont {{Rieffel}},\ and\ \citenamefont {{Wudarski}}}]{Levy_2022}%
  \BibitemOpen
  \bibfield  {author} {\bibinfo {author} {\bibfnamefont {R.}~\bibnamefont {{Levy}}}, \bibinfo {author} {\bibfnamefont {Z.}~\bibnamefont {{Gonzalez Izquierdo}}}, \bibinfo {author} {\bibfnamefont {Z.}~\bibnamefont {{Wang}}}, \bibinfo {author} {\bibfnamefont {J.}~\bibnamefont {{Marshall}}}, \bibinfo {author} {\bibfnamefont {J.}~\bibnamefont {{Barreto}}}, \bibinfo {author} {\bibfnamefont {L.}~\bibnamefont {{Fry-Bouriaux}}}, \bibinfo {author} {\bibfnamefont {D.~T.}\ \bibnamefont {{O'Connor}}}, \bibinfo {author} {\bibfnamefont {P.~A.}\ \bibnamefont {{Warburton}}}, \bibinfo {author} {\bibfnamefont {N.}~\bibnamefont {{Wiebe}}}, \bibinfo {author} {\bibfnamefont {E.}~\bibnamefont {{Rieffel}}},\ and\ \bibinfo {author} {\bibfnamefont {F.~A.}\ \bibnamefont {{Wudarski}}},\ }\bibfield  {title} {\bibinfo {title} {{Towards solving the Fermi-Hubbard model via tailored quantum annealers}},\ }\href {https://doi.org/10.48550/arXiv.2207.14374} {\bibfield  {journal} {\bibinfo  {journal} {arXiv e-prints}\ ,\ \bibinfo {eid}
  {arXiv:2207.14374}} (\bibinfo {year} {2022})},\ \Eprint {https://arxiv.org/abs/2207.14374} {arXiv:2207.14374 [quant-ph]} \BibitemShut {NoStop}%
\bibitem [{\citenamefont {Sud}\ \emph {et~al.}(2022)\citenamefont {Sud}, \citenamefont {Hadfield}, \citenamefont {Rieffel}, \citenamefont {Tubman},\ and\ \citenamefont {Hogg}}]{sud2022parameter}%
  \BibitemOpen
  \bibfield  {author} {\bibinfo {author} {\bibfnamefont {J.}~\bibnamefont {Sud}}, \bibinfo {author} {\bibfnamefont {S.}~\bibnamefont {Hadfield}}, \bibinfo {author} {\bibfnamefont {E.}~\bibnamefont {Rieffel}}, \bibinfo {author} {\bibfnamefont {N.}~\bibnamefont {Tubman}},\ and\ \bibinfo {author} {\bibfnamefont {T.}~\bibnamefont {Hogg}},\ }\href@noop {} {\bibinfo {title} {A parameter setting heuristic for the quantum alternating operator ansatz}} (\bibinfo {year} {2022}),\ \Eprint {https://arxiv.org/abs/2211.09270} {arXiv:2211.09270 [quant-ph]} \BibitemShut {NoStop}%
\bibitem [{\citenamefont {Streif}\ \emph {et~al.}(2021)\citenamefont {Streif}, \citenamefont {Leib}, \citenamefont {Wudarski}, \citenamefont {Rieffel},\ and\ \citenamefont {Wang}}]{Streif_2021}%
  \BibitemOpen
  \bibfield  {author} {\bibinfo {author} {\bibfnamefont {M.}~\bibnamefont {Streif}}, \bibinfo {author} {\bibfnamefont {M.}~\bibnamefont {Leib}}, \bibinfo {author} {\bibfnamefont {F.}~\bibnamefont {Wudarski}}, \bibinfo {author} {\bibfnamefont {E.}~\bibnamefont {Rieffel}},\ and\ \bibinfo {author} {\bibfnamefont {Z.}~\bibnamefont {Wang}},\ }\bibfield  {title} {\bibinfo {title} {Quantum algorithms with local particle-number conservation: Noise effects and error correction},\ }\bibfield  {journal} {\bibinfo  {journal} {Physical Review A}\ }\textbf {\bibinfo {volume} {103}},\ \href {https://doi.org/10.1103/physreva.103.042412} {10.1103/physreva.103.042412} (\bibinfo {year} {2021})\BibitemShut {NoStop}%
\bibitem [{\citenamefont {Kremenetski}\ \emph {et~al.}(2023)\citenamefont {Kremenetski}, \citenamefont {Apte}, \citenamefont {Hogg}, \citenamefont {Hadfield},\ and\ \citenamefont {Tubman}}]{kremenetski2023quantum}%
  \BibitemOpen
  \bibfield  {author} {\bibinfo {author} {\bibfnamefont {V.}~\bibnamefont {Kremenetski}}, \bibinfo {author} {\bibfnamefont {A.}~\bibnamefont {Apte}}, \bibinfo {author} {\bibfnamefont {T.}~\bibnamefont {Hogg}}, \bibinfo {author} {\bibfnamefont {S.}~\bibnamefont {Hadfield}},\ and\ \bibinfo {author} {\bibfnamefont {N.~M.}\ \bibnamefont {Tubman}},\ }\href@noop {} {\bibinfo {title} {Quantum alternating operator ansatz (qaoa) beyond low depth with gradually changing unitaries}} (\bibinfo {year} {2023}),\ \Eprint {https://arxiv.org/abs/2305.04455} {arXiv:2305.04455 [quant-ph]} \BibitemShut {NoStop}%
\bibitem [{\citenamefont {LaRose}\ \emph {et~al.}(2022)\citenamefont {LaRose}, \citenamefont {Rieffel},\ and\ \citenamefont {Venturelli}}]{LaRose_2022}%
  \BibitemOpen
  \bibfield  {author} {\bibinfo {author} {\bibfnamefont {R.}~\bibnamefont {LaRose}}, \bibinfo {author} {\bibfnamefont {E.}~\bibnamefont {Rieffel}},\ and\ \bibinfo {author} {\bibfnamefont {D.}~\bibnamefont {Venturelli}},\ }\bibfield  {title} {\bibinfo {title} {Mixer-phaser ansätze for quantum optimization with hard constraints},\ }\bibfield  {journal} {\bibinfo  {journal} {Quantum Machine Intelligence}\ }\textbf {\bibinfo {volume} {4}},\ \href {https://doi.org/10.1007/s42484-022-00069-x} {10.1007/s42484-022-00069-x} (\bibinfo {year} {2022})\BibitemShut {NoStop}%
\bibitem [{\citenamefont {Taassob}\ \emph {et~al.}(2023)\citenamefont {Taassob}, \citenamefont {Venturelli},\ and\ \citenamefont {Lott}}]{taassob2023}%
  \BibitemOpen
  \bibfield  {author} {\bibinfo {author} {\bibfnamefont {A.}~\bibnamefont {Taassob}}, \bibinfo {author} {\bibfnamefont {D.}~\bibnamefont {Venturelli}},\ and\ \bibinfo {author} {\bibfnamefont {P.~A.}\ \bibnamefont {Lott}},\ }\bibfield  {title} {\bibinfo {title} {Neural deep operator networks representation of coherent ising machine dynamics},\ }in\ \href {https://openreview.net/forum?id=3W3Qo3arAG} {\emph {\bibinfo {booktitle} {Machine Learning with New Compute Paradigms}}}\ (\bibinfo {year} {2023})\BibitemShut {NoStop}%
\bibitem [{\citenamefont {Gustafson}\ \emph {et~al.}(2023)\citenamefont {Gustafson}, \citenamefont {Li}, \citenamefont {Khan}, \citenamefont {Kim}, \citenamefont {Kurkcuoglu}, \citenamefont {Alam}, \citenamefont {Orth}, \citenamefont {Rahmani},\ and\ \citenamefont {Iadecola}}]{gustafson2023preparing}%
  \BibitemOpen
  \bibfield  {author} {\bibinfo {author} {\bibfnamefont {E.~J.}\ \bibnamefont {Gustafson}}, \bibinfo {author} {\bibfnamefont {A.~C.~Y.}\ \bibnamefont {Li}}, \bibinfo {author} {\bibfnamefont {A.}~\bibnamefont {Khan}}, \bibinfo {author} {\bibfnamefont {J.}~\bibnamefont {Kim}}, \bibinfo {author} {\bibfnamefont {D.~M.}\ \bibnamefont {Kurkcuoglu}}, \bibinfo {author} {\bibfnamefont {M.~S.}\ \bibnamefont {Alam}}, \bibinfo {author} {\bibfnamefont {P.~P.}\ \bibnamefont {Orth}}, \bibinfo {author} {\bibfnamefont {A.}~\bibnamefont {Rahmani}},\ and\ \bibinfo {author} {\bibfnamefont {T.}~\bibnamefont {Iadecola}},\ }\href@noop {} {\bibinfo {title} {Preparing quantum many-body scar states on quantum computers}} (\bibinfo {year} {2023}),\ \Eprint {https://arxiv.org/abs/2301.08226} {arXiv:2301.08226 [quant-ph]} \BibitemShut {NoStop}%
\bibitem [{\citenamefont {Hirsbrunner}\ \emph {et~al.}(2023)\citenamefont {Hirsbrunner}, \citenamefont {Chamaki}, \citenamefont {Mullinax},\ and\ \citenamefont {Tubman}}]{hirsbrunner2023mp2}%
  \BibitemOpen
  \bibfield  {author} {\bibinfo {author} {\bibfnamefont {M.~R.}\ \bibnamefont {Hirsbrunner}}, \bibinfo {author} {\bibfnamefont {D.}~\bibnamefont {Chamaki}}, \bibinfo {author} {\bibfnamefont {J.~W.}\ \bibnamefont {Mullinax}},\ and\ \bibinfo {author} {\bibfnamefont {N.~M.}\ \bibnamefont {Tubman}},\ }\href@noop {} {\bibinfo {title} {Beyond mp2 initialization for unitary coupled cluster quantum circuits}} (\bibinfo {year} {2023}),\ \Eprint {https://arxiv.org/abs/2301.05666} {arXiv:2301.05666 [quant-ph]} \BibitemShut {NoStop}%
\bibitem [{\citenamefont {Sorourifar}\ \emph {et~al.}(2024)\citenamefont {Sorourifar}, \citenamefont {Chamaki}, \citenamefont {Tubman}, \citenamefont {Paulson},\ and\ \citenamefont {{Bernal Neira}}}]{Sorourifar2024}%
  \BibitemOpen
  \bibfield  {author} {\bibinfo {author} {\bibfnamefont {F.}~\bibnamefont {Sorourifar}}, \bibinfo {author} {\bibfnamefont {D.}~\bibnamefont {Chamaki}}, \bibinfo {author} {\bibfnamefont {N.~M.}\ \bibnamefont {Tubman}}, \bibinfo {author} {\bibfnamefont {J.}~\bibnamefont {Paulson}},\ and\ \bibinfo {author} {\bibfnamefont {D.~E.}\ \bibnamefont {{Bernal Neira}}},\ }\bibfield  {title} {\bibinfo {title} {Bayesian optimization priors for efficient variational quantum algorithms},\ }in\ \href {https://doi.org/https://doi.org/10.1016/B978-0-443-28824-1.50564-0} {\emph {\bibinfo {booktitle} {34th European Symposium on Computer Aided Process Engineering / 15th International Symposium on Process Systems Engineering}}},\ \bibinfo {series} {Computer Aided Chemical Engineering}, Vol.~\bibinfo {volume} {53},\ \bibinfo {editor} {edited by\ \bibinfo {editor} {\bibfnamefont {F.}~\bibnamefont {Manenti}}\ and\ \bibinfo {editor} {\bibfnamefont {G.~V.}\ \bibnamefont {Reklaitis}}}\ (\bibinfo  {publisher} {Elsevier},\ \bibinfo {year}
  {2024})\ pp.\ \bibinfo {pages} {3379--3384}\BibitemShut {NoStop}%
\bibitem [{\citenamefont {Parolini}\ and\ \citenamefont {Mossi}(2020)}]{parolini2020multifractal}%
  \BibitemOpen
  \bibfield  {author} {\bibinfo {author} {\bibfnamefont {T.}~\bibnamefont {Parolini}}\ and\ \bibinfo {author} {\bibfnamefont {G.}~\bibnamefont {Mossi}},\ }\href@noop {} {\bibinfo {title} {Multifractal dynamics of the qrem}} (\bibinfo {year} {2020}),\ \Eprint {https://arxiv.org/abs/2007.00315} {arXiv:2007.00315 [cond-mat.dis-nn]} \BibitemShut {NoStop}%
\bibitem [{\citenamefont {Suri}\ \emph {et~al.}(2023)\citenamefont {Suri}, \citenamefont {Barreto}, \citenamefont {Hadfield}, \citenamefont {Wiebe}, \citenamefont {Wudarski},\ and\ \citenamefont {Marshall}}]{Suri_2023}%
  \BibitemOpen
  \bibfield  {author} {\bibinfo {author} {\bibfnamefont {N.}~\bibnamefont {Suri}}, \bibinfo {author} {\bibfnamefont {J.}~\bibnamefont {Barreto}}, \bibinfo {author} {\bibfnamefont {S.}~\bibnamefont {Hadfield}}, \bibinfo {author} {\bibfnamefont {N.}~\bibnamefont {Wiebe}}, \bibinfo {author} {\bibfnamefont {F.}~\bibnamefont {Wudarski}},\ and\ \bibinfo {author} {\bibfnamefont {J.}~\bibnamefont {Marshall}},\ }\bibfield  {title} {\bibinfo {title} {Two-{U}nitary {D}ecomposition {A}lgorithm and {O}pen {Q}uantum {S}ystem {S}imulation},\ }\href {https://doi.org/10.22331/q-2023-05-15-1002} {\bibfield  {journal} {\bibinfo  {journal} {{Quantum}}\ }\textbf {\bibinfo {volume} {7}},\ \bibinfo {pages} {1002} (\bibinfo {year} {2023})}\BibitemShut {NoStop}%
\bibitem [{\citenamefont {Chamaki}\ \emph {et~al.}(2022)\citenamefont {Chamaki}, \citenamefont {Hadfield}, \citenamefont {Klymko}, \citenamefont {O'Gorman},\ and\ \citenamefont {Tubman}}]{chamaki2022selfconsistent}%
  \BibitemOpen
  \bibfield  {author} {\bibinfo {author} {\bibfnamefont {D.~B.}\ \bibnamefont {Chamaki}}, \bibinfo {author} {\bibfnamefont {S.}~\bibnamefont {Hadfield}}, \bibinfo {author} {\bibfnamefont {K.}~\bibnamefont {Klymko}}, \bibinfo {author} {\bibfnamefont {B.}~\bibnamefont {O'Gorman}},\ and\ \bibinfo {author} {\bibfnamefont {N.~M.}\ \bibnamefont {Tubman}},\ }\href@noop {} {\bibinfo {title} {Self-consistent quantum iteratively sparsified hamiltonian method (squish): A new algorithm for efficient hamiltonian simulation and compression}} (\bibinfo {year} {2022}),\ \Eprint {https://arxiv.org/abs/2211.16522} {arXiv:2211.16522 [quant-ph]} \BibitemShut {NoStop}%
\bibitem [{\citenamefont {Verdon}\ \emph {et~al.}(2019)\citenamefont {Verdon}, \citenamefont {Marks}, \citenamefont {Nanda}, \citenamefont {Leichenauer},\ and\ \citenamefont {Hidary}}]{verdon2019quantum}%
  \BibitemOpen
  \bibfield  {author} {\bibinfo {author} {\bibfnamefont {G.}~\bibnamefont {Verdon}}, \bibinfo {author} {\bibfnamefont {J.}~\bibnamefont {Marks}}, \bibinfo {author} {\bibfnamefont {S.}~\bibnamefont {Nanda}}, \bibinfo {author} {\bibfnamefont {S.}~\bibnamefont {Leichenauer}},\ and\ \bibinfo {author} {\bibfnamefont {J.}~\bibnamefont {Hidary}},\ }\href@noop {} {\bibinfo {title} {Quantum hamiltonian-based models and the variational quantum thermalizer algorithm}} (\bibinfo {year} {2019}),\ \Eprint {https://arxiv.org/abs/1910.02071} {arXiv:1910.02071 [quant-ph]} \BibitemShut {NoStop}%
\bibitem [{\citenamefont {Templin}\ \emph {et~al.}(2023)\citenamefont {Templin}, \citenamefont {Memarzadeh}, \citenamefont {Vinci}, \citenamefont {Lott}, \citenamefont {Akbari~Asanjan}, \citenamefont {Alexiades~Armenakas},\ and\ \citenamefont {Rieffel}}]{Templin2023}%
  \BibitemOpen
  \bibfield  {author} {\bibinfo {author} {\bibfnamefont {T.}~\bibnamefont {Templin}}, \bibinfo {author} {\bibfnamefont {M.}~\bibnamefont {Memarzadeh}}, \bibinfo {author} {\bibfnamefont {W.}~\bibnamefont {Vinci}}, \bibinfo {author} {\bibfnamefont {P.~A.}\ \bibnamefont {Lott}}, \bibinfo {author} {\bibfnamefont {A.}~\bibnamefont {Akbari~Asanjan}}, \bibinfo {author} {\bibfnamefont {A.}~\bibnamefont {Alexiades~Armenakas}},\ and\ \bibinfo {author} {\bibfnamefont {E.}~\bibnamefont {Rieffel}},\ }\bibfield  {title} {\bibinfo {title} {Anomaly detection in aeronautics data with quantum-compatible discrete deep generative model},\ }\href {https://doi.org/10.1088/2632-2153/ace756} {\bibfield  {journal} {\bibinfo  {journal} {Machine Learning: Science and Technology}\ }\textbf {\bibinfo {volume} {4}},\ \bibinfo {pages} {035018} (\bibinfo {year} {2023})}\BibitemShut {NoStop}%
\bibitem [{\citenamefont {Brown}\ \emph {et~al.}(2023)\citenamefont {Brown}, \citenamefont {Neira}, \citenamefont {Venturelli},\ and\ \citenamefont {Pavone}}]{brown2023copositive}%
  \BibitemOpen
  \bibfield  {author} {\bibinfo {author} {\bibfnamefont {R.}~\bibnamefont {Brown}}, \bibinfo {author} {\bibfnamefont {D.~E.~B.}\ \bibnamefont {Neira}}, \bibinfo {author} {\bibfnamefont {D.}~\bibnamefont {Venturelli}},\ and\ \bibinfo {author} {\bibfnamefont {M.}~\bibnamefont {Pavone}},\ }\href@noop {} {\bibinfo {title} {A copositive framework for analysis of hybrid ising-classical algorithms}} (\bibinfo {year} {2023}),\ \Eprint {https://arxiv.org/abs/2207.13630} {arXiv:2207.13630 [math.OC]} \BibitemShut {NoStop}%
\bibitem [{\citenamefont {{Khan}}\ \emph {et~al.}(2024)\citenamefont {{Khan}}, \citenamefont {{Vaish}}, \citenamefont {{Pang}}, \citenamefont {{Kowshik}}, \citenamefont {{Chen}}, \citenamefont {{Batton}}, \citenamefont {{Rotskoff}}, \citenamefont {{Mullinax}}, \citenamefont {{Clark}}, \citenamefont {{Rubenstein}},\ and\ \citenamefont {{Tubman}}}]{Khan2024}%
  \BibitemOpen
  \bibfield  {author} {\bibinfo {author} {\bibfnamefont {A.}~\bibnamefont {{Khan}}}, \bibinfo {author} {\bibfnamefont {P.}~\bibnamefont {{Vaish}}}, \bibinfo {author} {\bibfnamefont {Y.}~\bibnamefont {{Pang}}}, \bibinfo {author} {\bibfnamefont {N.}~\bibnamefont {{Kowshik}}}, \bibinfo {author} {\bibfnamefont {M.~S.}\ \bibnamefont {{Chen}}}, \bibinfo {author} {\bibfnamefont {C.~H.}\ \bibnamefont {{Batton}}}, \bibinfo {author} {\bibfnamefont {G.~M.}\ \bibnamefont {{Rotskoff}}}, \bibinfo {author} {\bibfnamefont {J.~W.}\ \bibnamefont {{Mullinax}}}, \bibinfo {author} {\bibfnamefont {B.~K.}\ \bibnamefont {{Clark}}}, \bibinfo {author} {\bibfnamefont {B.~M.}\ \bibnamefont {{Rubenstein}}},\ and\ \bibinfo {author} {\bibfnamefont {N.~M.}\ \bibnamefont {{Tubman}}},\ }\bibfield  {title} {\bibinfo {title} {{Quantum Hardware-Enabled Molecular Dynamics via Transfer Learning}},\ }\href {https://doi.org/10.48550/arXiv.2406.08554} {\bibfield  {journal} {\bibinfo  {journal} {arXiv e-prints}\ ,\ \bibinfo {eid} {arXiv:2406.08554}}
  (\bibinfo {year} {2024})},\ \Eprint {https://arxiv.org/abs/2406.08554} {arXiv:2406.08554 [physics.chem-ph]} \BibitemShut {NoStop}%
\bibitem [{\citenamefont {{Khan}}\ \emph {et~al.}(2023)\citenamefont {{Khan}}, \citenamefont {{Clark}},\ and\ \citenamefont {{Tubman}}}]{Khan2023}%
  \BibitemOpen
  \bibfield  {author} {\bibinfo {author} {\bibfnamefont {A.}~\bibnamefont {{Khan}}}, \bibinfo {author} {\bibfnamefont {B.~K.}\ \bibnamefont {{Clark}}},\ and\ \bibinfo {author} {\bibfnamefont {N.~M.}\ \bibnamefont {{Tubman}}},\ }\bibfield  {title} {\bibinfo {title} {{Pre-optimizing variational quantum eigensolvers with tensor networks}},\ }\href {https://doi.org/10.48550/arXiv.2310.12965} {\bibfield  {journal} {\bibinfo  {journal} {arXiv e-prints}\ ,\ \bibinfo {eid} {arXiv:2310.12965}} (\bibinfo {year} {2023})},\ \Eprint {https://arxiv.org/abs/2310.12965} {arXiv:2310.12965 [quant-ph]} \BibitemShut {NoStop}%
\bibitem [{\citenamefont {{Gustafson}}\ \emph {et~al.}(2024)\citenamefont {{Gustafson}}, \citenamefont {{Tiihonen}}, \citenamefont {{Chamaki}}, \citenamefont {{Sorourifar}}, \citenamefont {{Mullinax}}, \citenamefont {{Li}}, \citenamefont {{Maciejewski}}, \citenamefont {{Sawaya}}, \citenamefont {{Krogel}}, \citenamefont {{Bernal Neira}},\ and\ \citenamefont {{Tubman}}}]{Gustafson2024}%
  \BibitemOpen
  \bibfield  {author} {\bibinfo {author} {\bibfnamefont {E.~J.}\ \bibnamefont {{Gustafson}}}, \bibinfo {author} {\bibfnamefont {J.}~\bibnamefont {{Tiihonen}}}, \bibinfo {author} {\bibfnamefont {D.}~\bibnamefont {{Chamaki}}}, \bibinfo {author} {\bibfnamefont {F.}~\bibnamefont {{Sorourifar}}}, \bibinfo {author} {\bibfnamefont {J.~W.}\ \bibnamefont {{Mullinax}}}, \bibinfo {author} {\bibfnamefont {A.~C.~Y.}\ \bibnamefont {{Li}}}, \bibinfo {author} {\bibfnamefont {F.~B.}\ \bibnamefont {{Maciejewski}}}, \bibinfo {author} {\bibfnamefont {N.~P.}\ \bibnamefont {{Sawaya}}}, \bibinfo {author} {\bibfnamefont {J.~T.}\ \bibnamefont {{Krogel}}}, \bibinfo {author} {\bibfnamefont {D.~E.}\ \bibnamefont {{Bernal Neira}}},\ and\ \bibinfo {author} {\bibfnamefont {N.~M.}\ \bibnamefont {{Tubman}}},\ }\bibfield  {title} {\bibinfo {title} {{Surrogate optimization of variational quantum circuits}},\ }\href {https://doi.org/10.48550/arXiv.2404.02951} {\bibfield  {journal} {\bibinfo  {journal} {arXiv e-prints}\ ,\ \bibinfo {eid}
  {arXiv:2404.02951}} (\bibinfo {year} {2024})},\ \Eprint {https://arxiv.org/abs/2404.02951} {arXiv:2404.02951 [quant-ph]} \BibitemShut {NoStop}%
\bibitem [{\citenamefont {Broughton}\ \emph {et~al.}(2021)\citenamefont {Broughton}, \citenamefont {Verdon}, \citenamefont {McCourt}, \citenamefont {Martinez}, \citenamefont {Yoo}, \citenamefont {Isakov}, \citenamefont {Massey}, \citenamefont {Halavati}, \citenamefont {Niu}, \citenamefont {Zlokapa}, \citenamefont {Peters}, \citenamefont {Lockwood}, \citenamefont {Skolik}, \citenamefont {Jerbi}, \citenamefont {Dunjko}, \citenamefont {Leib}, \citenamefont {Streif}, \citenamefont {Dollen}, \citenamefont {Chen}, \citenamefont {Cao}, \citenamefont {Wiersema}, \citenamefont {Huang}, \citenamefont {McClean}, \citenamefont {Babbush}, \citenamefont {Boixo}, \citenamefont {Bacon}, \citenamefont {Ho}, \citenamefont {Neven},\ and\ \citenamefont {Mohseni}}]{broughton2021tensorflow}%
  \BibitemOpen
  \bibfield  {author} {\bibinfo {author} {\bibfnamefont {M.}~\bibnamefont {Broughton}}, \bibinfo {author} {\bibfnamefont {G.}~\bibnamefont {Verdon}}, \bibinfo {author} {\bibfnamefont {T.}~\bibnamefont {McCourt}}, \bibinfo {author} {\bibfnamefont {A.~J.}\ \bibnamefont {Martinez}}, \bibinfo {author} {\bibfnamefont {J.~H.}\ \bibnamefont {Yoo}}, \bibinfo {author} {\bibfnamefont {S.~V.}\ \bibnamefont {Isakov}}, \bibinfo {author} {\bibfnamefont {P.}~\bibnamefont {Massey}}, \bibinfo {author} {\bibfnamefont {R.}~\bibnamefont {Halavati}}, \bibinfo {author} {\bibfnamefont {M.~Y.}\ \bibnamefont {Niu}}, \bibinfo {author} {\bibfnamefont {A.}~\bibnamefont {Zlokapa}}, \bibinfo {author} {\bibfnamefont {E.}~\bibnamefont {Peters}}, \bibinfo {author} {\bibfnamefont {O.}~\bibnamefont {Lockwood}}, \bibinfo {author} {\bibfnamefont {A.}~\bibnamefont {Skolik}}, \bibinfo {author} {\bibfnamefont {S.}~\bibnamefont {Jerbi}}, \bibinfo {author} {\bibfnamefont {V.}~\bibnamefont {Dunjko}}, \bibinfo {author} {\bibfnamefont {M.}~\bibnamefont
  {Leib}}, \bibinfo {author} {\bibfnamefont {M.}~\bibnamefont {Streif}}, \bibinfo {author} {\bibfnamefont {D.~V.}\ \bibnamefont {Dollen}}, \bibinfo {author} {\bibfnamefont {H.}~\bibnamefont {Chen}}, \bibinfo {author} {\bibfnamefont {S.}~\bibnamefont {Cao}}, \bibinfo {author} {\bibfnamefont {R.}~\bibnamefont {Wiersema}}, \bibinfo {author} {\bibfnamefont {H.-Y.}\ \bibnamefont {Huang}}, \bibinfo {author} {\bibfnamefont {J.~R.}\ \bibnamefont {McClean}}, \bibinfo {author} {\bibfnamefont {R.}~\bibnamefont {Babbush}}, \bibinfo {author} {\bibfnamefont {S.}~\bibnamefont {Boixo}}, \bibinfo {author} {\bibfnamefont {D.}~\bibnamefont {Bacon}}, \bibinfo {author} {\bibfnamefont {A.~K.}\ \bibnamefont {Ho}}, \bibinfo {author} {\bibfnamefont {H.}~\bibnamefont {Neven}},\ and\ \bibinfo {author} {\bibfnamefont {M.}~\bibnamefont {Mohseni}},\ }\href@noop {} {\bibinfo {title} {Tensorflow quantum: A software framework for quantum machine learning}} (\bibinfo {year} {2021}),\ \Eprint {https://arxiv.org/abs/2003.02989}
  {arXiv:2003.02989 [quant-ph]} \BibitemShut {NoStop}%
\bibitem [{\citenamefont {Villalonga}\ \emph {et~al.}(2019)\citenamefont {Villalonga}, \citenamefont {Boixo}, \citenamefont {Nelson}, \citenamefont {Henze}, \citenamefont {Rieffel}, \citenamefont {Biswas},\ and\ \citenamefont {Mandr{\`{a}}}}]{Villalonga2019}%
  \BibitemOpen
  \bibfield  {author} {\bibinfo {author} {\bibfnamefont {B.}~\bibnamefont {Villalonga}}, \bibinfo {author} {\bibfnamefont {S.}~\bibnamefont {Boixo}}, \bibinfo {author} {\bibfnamefont {B.}~\bibnamefont {Nelson}}, \bibinfo {author} {\bibfnamefont {C.}~\bibnamefont {Henze}}, \bibinfo {author} {\bibfnamefont {E.}~\bibnamefont {Rieffel}}, \bibinfo {author} {\bibfnamefont {R.}~\bibnamefont {Biswas}},\ and\ \bibinfo {author} {\bibfnamefont {S.}~\bibnamefont {Mandr{\`{a}}}},\ }\bibfield  {title} {\bibinfo {title} {A flexible high-performance simulator for verifying and benchmarking quantum circuits implemented on real hardware},\ }\bibfield  {journal} {\bibinfo  {journal} {npj Quantum Information}\ }\textbf {\bibinfo {volume} {5}},\ \href {https://doi.org/10.1038/s41534-019-0196-1} {10.1038/s41534-019-0196-1} (\bibinfo {year} {2019})\BibitemShut {NoStop}%
\bibitem [{\citenamefont {Chowdhury}\ \emph {et~al.}(2025)\citenamefont {Chowdhury}, \citenamefont {Aadit}, \citenamefont {Grimaldi}, \citenamefont {Raimondo}, \citenamefont {Raut}, \citenamefont {Lott}, \citenamefont {Mentink}, \citenamefont {Rams}, \citenamefont {Ricci-Tersenghi}, \citenamefont {Chiappini}, \citenamefont {Theogarajan}, \citenamefont {Srimani}, \citenamefont {Finocchio}, \citenamefont {Mohseni},\ and\ \citenamefont {Camsari}}]{chowdhury2025pushingboundaryquantumadvantage}%
  \BibitemOpen
  \bibfield  {author} {\bibinfo {author} {\bibfnamefont {S.}~\bibnamefont {Chowdhury}}, \bibinfo {author} {\bibfnamefont {N.~A.}\ \bibnamefont {Aadit}}, \bibinfo {author} {\bibfnamefont {A.}~\bibnamefont {Grimaldi}}, \bibinfo {author} {\bibfnamefont {E.}~\bibnamefont {Raimondo}}, \bibinfo {author} {\bibfnamefont {A.}~\bibnamefont {Raut}}, \bibinfo {author} {\bibfnamefont {P.~A.}\ \bibnamefont {Lott}}, \bibinfo {author} {\bibfnamefont {J.~H.}\ \bibnamefont {Mentink}}, \bibinfo {author} {\bibfnamefont {M.~M.}\ \bibnamefont {Rams}}, \bibinfo {author} {\bibfnamefont {F.}~\bibnamefont {Ricci-Tersenghi}}, \bibinfo {author} {\bibfnamefont {M.}~\bibnamefont {Chiappini}}, \bibinfo {author} {\bibfnamefont {L.~S.}\ \bibnamefont {Theogarajan}}, \bibinfo {author} {\bibfnamefont {T.}~\bibnamefont {Srimani}}, \bibinfo {author} {\bibfnamefont {G.}~\bibnamefont {Finocchio}}, \bibinfo {author} {\bibfnamefont {M.}~\bibnamefont {Mohseni}},\ and\ \bibinfo {author} {\bibfnamefont {K.~Y.}\ \bibnamefont {Camsari}},\ }\href
  {https://arxiv.org/abs/2503.10302} {\bibinfo {title} {Pushing the boundary of quantum advantage in hard combinatorial optimization with probabilistic computers}} (\bibinfo {year} {2025}),\ \Eprint {https://arxiv.org/abs/2503.10302} {arXiv:2503.10302 [quant-ph]} \BibitemShut {NoStop}%
\bibitem [{\citenamefont {Aadit}\ \emph {et~al.}(2023{\natexlab{a}})\citenamefont {Aadit}, \citenamefont {Lott},\ and\ \citenamefont {Mohseni}}]{Aadit_Nonlocal_Monte_Carlor_2023}%
  \BibitemOpen
  \bibfield  {author} {\bibinfo {author} {\bibfnamefont {N.}~\bibnamefont {Aadit}}, \bibinfo {author} {\bibfnamefont {P.~A.}\ \bibnamefont {Lott}},\ and\ \bibinfo {author} {\bibfnamefont {M.}~\bibnamefont {Mohseni}},\ }\href {https://github.com/usra-riacs/Nonlocal-Monte-Carlo} {\bibinfo {title} {{Nonlocal Monte Carlor}}} (\bibinfo {year} {2023}{\natexlab{a}})\BibitemShut {NoStop}%
\bibitem [{\citenamefont {Aadit}\ \emph {et~al.}(2023{\natexlab{b}})\citenamefont {Aadit}, \citenamefont {Lott},\ and\ \citenamefont {Mohseni}}]{Aadit_APT-solver_An_Adaptive_2023}%
  \BibitemOpen
  \bibfield  {author} {\bibinfo {author} {\bibfnamefont {N.}~\bibnamefont {Aadit}}, \bibinfo {author} {\bibfnamefont {P.~A.}\ \bibnamefont {Lott}},\ and\ \bibinfo {author} {\bibfnamefont {M.}~\bibnamefont {Mohseni}},\ }\href {https://github.com/usra-riacs/APT-solver} {\bibinfo {title} {{APT-solver: An Adaptive Parallel Tempering Solver}}} (\bibinfo {year} {2023}{\natexlab{b}})\BibitemShut {NoStop}%
\bibitem [{\citenamefont {Bernal~Neira}\ \emph {et~al.}(2023)\citenamefont {Bernal~Neira}, \citenamefont {Brown}, \citenamefont {Sathe},\ and\ \citenamefont {Venturelli}}]{Bernal_Neira_Stochastic_Benchmark_toolkit_2023}%
  \BibitemOpen
  \bibfield  {author} {\bibinfo {author} {\bibfnamefont {D.~E.}\ \bibnamefont {Bernal~Neira}}, \bibinfo {author} {\bibfnamefont {R.}~\bibnamefont {Brown}}, \bibinfo {author} {\bibfnamefont {P.}~\bibnamefont {Sathe}},\ and\ \bibinfo {author} {\bibfnamefont {D.}~\bibnamefont {Venturelli}},\ }\href {https://github.com/usra-riacs/stochastic-benchmark} {\bibinfo {title} {{Stochastic Benchmark: toolkit for performance evaluation and parameter tuning of stochastic parameterized stochastic optimization solvers}}} (\bibinfo {year} {2023})\BibitemShut {NoStop}%
\bibitem [{\citenamefont {{Bernal Neira}}\ \emph {et~al.}(2024)\citenamefont {{Bernal Neira}}, \citenamefont {{Brown}}, \citenamefont {{Sathe}}, \citenamefont {{Wudarski}}, \citenamefont {{Pavone}}, \citenamefont {{Rieffel}},\ and\ \citenamefont {{Venturelli}}}]{Bernal2024}%
  \BibitemOpen
  \bibfield  {author} {\bibinfo {author} {\bibfnamefont {D.~E.}\ \bibnamefont {{Bernal Neira}}}, \bibinfo {author} {\bibfnamefont {R.}~\bibnamefont {{Brown}}}, \bibinfo {author} {\bibfnamefont {P.}~\bibnamefont {{Sathe}}}, \bibinfo {author} {\bibfnamefont {F.}~\bibnamefont {{Wudarski}}}, \bibinfo {author} {\bibfnamefont {M.}~\bibnamefont {{Pavone}}}, \bibinfo {author} {\bibfnamefont {E.~G.}\ \bibnamefont {{Rieffel}}},\ and\ \bibinfo {author} {\bibfnamefont {D.}~\bibnamefont {{Venturelli}}},\ }\bibfield  {title} {\bibinfo {title} {{Benchmarking the Operation of Quantum Heuristics and Ising Machines: Scoring Parameter Setting Strategies on Optimization Applications}},\ }\href {https://doi.org/10.48550/arXiv.2402.10255} {\bibfield  {journal} {\bibinfo  {journal} {arXiv e-prints}\ ,\ \bibinfo {eid} {arXiv:2402.10255}} (\bibinfo {year} {2024})},\ \Eprint {https://arxiv.org/abs/2402.10255} {arXiv:2402.10255 [quant-ph]} \BibitemShut {NoStop}%
\bibitem [{\citenamefont {Claes}\ \emph {et~al.}(2021)\citenamefont {Claes}, \citenamefont {Rieffel},\ and\ \citenamefont {Wang}}]{Claes_2021}%
  \BibitemOpen
  \bibfield  {author} {\bibinfo {author} {\bibfnamefont {J.}~\bibnamefont {Claes}}, \bibinfo {author} {\bibfnamefont {E.}~\bibnamefont {Rieffel}},\ and\ \bibinfo {author} {\bibfnamefont {Z.}~\bibnamefont {Wang}},\ }\bibfield  {title} {\bibinfo {title} {Character randomized benchmarking for non-multiplicity-free groups with applications to subspace, leakage, and matchgate randomized benchmarking},\ }\href {https://doi.org/10.1103/PRXQuantum.2.010351} {\bibfield  {journal} {\bibinfo  {journal} {PRX Quantum}\ }\textbf {\bibinfo {volume} {2}},\ \bibinfo {pages} {010351} (\bibinfo {year} {2021})}\BibitemShut {NoStop}%
\bibitem [{\citenamefont {Hu}\ \emph {et~al.}(2022)\citenamefont {Hu}, \citenamefont {LaRose}, \citenamefont {You}, \citenamefont {Rieffel},\ and\ \citenamefont {Wang}}]{hu2022logical}%
  \BibitemOpen
  \bibfield  {author} {\bibinfo {author} {\bibfnamefont {H.-Y.}\ \bibnamefont {Hu}}, \bibinfo {author} {\bibfnamefont {R.}~\bibnamefont {LaRose}}, \bibinfo {author} {\bibfnamefont {Y.-Z.}\ \bibnamefont {You}}, \bibinfo {author} {\bibfnamefont {E.}~\bibnamefont {Rieffel}},\ and\ \bibinfo {author} {\bibfnamefont {Z.}~\bibnamefont {Wang}},\ }\href@noop {} {\bibinfo {title} {Logical shadow tomography: Efficient estimation of error-mitigated observables}} (\bibinfo {year} {2022}),\ \Eprint {https://arxiv.org/abs/2203.07263} {arXiv:2203.07263 [quant-ph]} \BibitemShut {NoStop}%
\bibitem [{\citenamefont {Lubinski}\ \emph {et~al.}(2024)\citenamefont {Lubinski}, \citenamefont {Coffrin}, \citenamefont {McGeoch}, \citenamefont {Sathe}, \citenamefont {Apanavicius}, \citenamefont {Bernal~Neira},\ and\ \citenamefont {Collaboration}}]{lubinski2024}%
  \BibitemOpen
  \bibfield  {author} {\bibinfo {author} {\bibfnamefont {T.}~\bibnamefont {Lubinski}}, \bibinfo {author} {\bibfnamefont {C.}~\bibnamefont {Coffrin}}, \bibinfo {author} {\bibfnamefont {C.}~\bibnamefont {McGeoch}}, \bibinfo {author} {\bibfnamefont {P.}~\bibnamefont {Sathe}}, \bibinfo {author} {\bibfnamefont {J.}~\bibnamefont {Apanavicius}}, \bibinfo {author} {\bibfnamefont {D.}~\bibnamefont {Bernal~Neira}},\ and\ \bibinfo {author} {\bibfnamefont {Q.~E. D. C.-C.}\ \bibnamefont {Collaboration}},\ }\bibfield  {title} {\bibinfo {title} {Optimization applications as quantum performance benchmarks},\ }\bibfield  {journal} {\bibinfo  {journal} {ACM Transactions on Quantum Computing}\ }\textbf {\bibinfo {volume} {5}},\ \href {https://doi.org/10.1145/3678184} {10.1145/3678184} (\bibinfo {year} {2024})\BibitemShut {NoStop}%
\bibitem [{\citenamefont {Leipold}\ \emph {et~al.}(2022)\citenamefont {Leipold}, \citenamefont {Spedalieri},\ and\ \citenamefont {Rieffel}}]{a15100356}%
  \BibitemOpen
  \bibfield  {author} {\bibinfo {author} {\bibfnamefont {H.}~\bibnamefont {Leipold}}, \bibinfo {author} {\bibfnamefont {F.~M.}\ \bibnamefont {Spedalieri}},\ and\ \bibinfo {author} {\bibfnamefont {E.}~\bibnamefont {Rieffel}},\ }\bibfield  {title} {\bibinfo {title} {Tailored quantum alternating operator ansatzes for circuit fault diagnostics},\ }\bibfield  {journal} {\bibinfo  {journal} {Algorithms}\ }\textbf {\bibinfo {volume} {15}},\ \href {https://doi.org/10.3390/a15100356} {10.3390/a15100356} (\bibinfo {year} {2022})\BibitemShut {NoStop}%
\bibitem [{\citenamefont {Gao}\ \emph {et~al.}(2020)\citenamefont {Gao}, \citenamefont {Wilson}, \citenamefont {Vandal}, \citenamefont {Vinci}, \citenamefont {Nemani},\ and\ \citenamefont {Rieffel}}]{Gao2020}%
  \BibitemOpen
  \bibfield  {author} {\bibinfo {author} {\bibfnamefont {N.}~\bibnamefont {Gao}}, \bibinfo {author} {\bibfnamefont {M.}~\bibnamefont {Wilson}}, \bibinfo {author} {\bibfnamefont {T.}~\bibnamefont {Vandal}}, \bibinfo {author} {\bibfnamefont {W.}~\bibnamefont {Vinci}}, \bibinfo {author} {\bibfnamefont {R.}~\bibnamefont {Nemani}},\ and\ \bibinfo {author} {\bibfnamefont {E.}~\bibnamefont {Rieffel}},\ }\bibfield  {title} {\bibinfo {title} {High-dimensional similarity search with quantum-assisted variational autoencoder},\ }in\ \href {https://doi.org/10.1145/3394486.3403138} {\emph {\bibinfo {booktitle} {Proceedings of the 26th ACM SIGKDD International Conference on Knowledge Discovery \& Data Mining}}},\ \bibinfo {series and number} {KDD '20}\ (\bibinfo  {publisher} {Association for Computing Machinery},\ \bibinfo {address} {New York, NY, USA},\ \bibinfo {year} {2020})\ p.\ \bibinfo {pages} {956–964}\BibitemShut {NoStop}%
\bibitem [{\citenamefont {Wilson}\ \emph {et~al.}(2021{\natexlab{a}})\citenamefont {Wilson}, \citenamefont {Vandal},\ and\ \citenamefont {Hogg}}]{Wilson2021}%
  \BibitemOpen
  \bibfield  {author} {\bibinfo {author} {\bibfnamefont {M.}~\bibnamefont {Wilson}}, \bibinfo {author} {\bibfnamefont {T.}~\bibnamefont {Vandal}},\ and\ \bibinfo {author} {\bibfnamefont {T.~e.~a.}\ \bibnamefont {Hogg}},\ }\bibfield  {title} {\bibinfo {title} {Quantum-assisted associative adversarial network: applying quantum annealing in deep learning},\ }\bibfield  {journal} {\bibinfo  {journal} {Quantum Machine Intelligence}\ }\textbf {\bibinfo {volume} {3}},\ \href {https://doi.org/10.1007/s42484-021-00047-9} {10.1007/s42484-021-00047-9} (\bibinfo {year} {2021}{\natexlab{a}})\BibitemShut {NoStop}%
\bibitem [{\citenamefont {Wilson}\ \emph {et~al.}(2021{\natexlab{b}})\citenamefont {Wilson}, \citenamefont {Stromswold},\ and\ \citenamefont {Wudarski}}]{Wilson2021b}%
  \BibitemOpen
  \bibfield  {author} {\bibinfo {author} {\bibfnamefont {M.}~\bibnamefont {Wilson}}, \bibinfo {author} {\bibfnamefont {R.}~\bibnamefont {Stromswold}},\ and\ \bibinfo {author} {\bibfnamefont {F.~e.~a.}\ \bibnamefont {Wudarski}},\ }\bibfield  {title} {\bibinfo {title} {Optimizing quantum heuristics with meta-learning},\ }\bibfield  {journal} {\bibinfo  {journal} {Quantum Machine Intelligence}\ }\textbf {\bibinfo {volume} {3}},\ \href {https://doi.org/10.1007/s42484-020-00022-w} {10.1007/s42484-020-00022-w} (\bibinfo {year} {2021}{\natexlab{b}})\BibitemShut {NoStop}%
\bibitem [{\citenamefont {Akbari~Asanjan}\ \emph {et~al.}(2023)\citenamefont {Akbari~Asanjan}, \citenamefont {Memarzadeh}, \citenamefont {Lott}, \citenamefont {Rieffel},\ and\ \citenamefont {Grabbe}}]{Asanjan2023}%
  \BibitemOpen
  \bibfield  {author} {\bibinfo {author} {\bibfnamefont {A.}~\bibnamefont {Akbari~Asanjan}}, \bibinfo {author} {\bibfnamefont {M.}~\bibnamefont {Memarzadeh}}, \bibinfo {author} {\bibfnamefont {P.~A.}\ \bibnamefont {Lott}}, \bibinfo {author} {\bibfnamefont {E.}~\bibnamefont {Rieffel}},\ and\ \bibinfo {author} {\bibfnamefont {S.}~\bibnamefont {Grabbe}},\ }\bibfield  {title} {\bibinfo {title} {Probabilistic wildfire segmentation using supervised deep generative model from satellite imagery},\ }\bibfield  {journal} {\bibinfo  {journal} {Remote Sensing}\ }\textbf {\bibinfo {volume} {15}},\ \href {https://doi.org/10.3390/rs15112718} {10.3390/rs15112718} (\bibinfo {year} {2023})\BibitemShut {NoStop}%
\bibitem [{\citenamefont {{O'Connor}}\ and\ \citenamefont {{Vinci}}(2020)}]{oconnor2021}%
  \BibitemOpen
  \bibfield  {author} {\bibinfo {author} {\bibfnamefont {D.}~\bibnamefont {{O'Connor}}}\ and\ \bibinfo {author} {\bibfnamefont {W.}~\bibnamefont {{Vinci}}},\ }\bibfield  {title} {\bibinfo {title} {{RBM-Flow and D-Flow: Invertible Flows with Discrete Energy Base Spaces}},\ }\href {https://doi.org/10.48550/arXiv.2012.13196} {\bibfield  {journal} {\bibinfo  {journal} {arXiv e-prints}\ ,\ \bibinfo {eid} {arXiv:2012.13196}} (\bibinfo {year} {2020})},\ \Eprint {https://arxiv.org/abs/2012.13196} {arXiv:2012.13196 [cs.LG]} \BibitemShut {NoStop}%
\end{thebibliography}%

\appendix

\section{Research Areas and Output of the Feynman Quantum Academy}
\label{research}

\subsection{Quantum Computing Architectures and Experimental Designs}
\subsubsection{Programmable Quantum Processor Applications}

A study focused on the dihedral group $D_N$ as an approximation for $U(1) \times \mathbb{Z}_2$ lattice gauge theory, showcased the development of efficient quantum circuits for key operations like the non-abelian Fourier transform, trace operation, and group multiplication and inversion \cite{Alam_2022}. The research emphasized that the required quantum resources for these operations scale linearly or as low-degree polynomials in $n = log(N)$, indicating a significant efficiency advantage over classical methods. 
 
These gates were successfully benchmarked on the Rigetti Aspen-9 quantum processor for $D_4$, achieving estimated fidelities exceeding 80\%. This high success rate underscores the potential of these quantum circuits in enabling large-scale lattice simulations of gauge theories. The paper's forward-looking perspective suggests future research directions, including extending these gates to more complex gauge theories and performing detailed resource analysis on specific architectures, promising advancements in quantum computing.  
 
The practical applications of programmable quantum processors have also been investigated. Demonstrated on IBM’s quantum processor for the Transverse-Field Ising Model (TFIM), an innovative algorithm using the Jarzynski equality to approximate free energy differences in quantum systems showcased the potential of quantum computers in efficiently simulating complex quantum systems, a task challenging for classical computers \cite{bassman2021computing}. This research complements the previous study's focus on the simulation of gauge theories, expanding the scope of quantum computational applications to include critical thermodynamic properties, thereby broadening the realm of quantum simulations. 

Another interesting application stems from the relationship between quantum annealing (QA) and Boltzmann Machines (BMs), as both QA and BMs are closely connected to the Boltzmann distribution. BMs are a type of generative artificial neural network that aim to learn the distribution of some training data set by fitting a Boltzmann distribution to the data. On the other hand, QA aims to produce approximate minimum energy (maximum likelihood) solutions to a Boltzmann distribution via finding the ground state of the associated Hamiltonian that determines the distribution. Taking advantage of this connection,~\cite{Kerger2023} proposes a framework for binary image denoising via restricted Boltzmann machines (RBMs) that introduces a denoising objective in quadratic unconstrained binary optimization (QUBO) form and is thus well-suited for QA.

The task of image denoising is a fundamental problem in image processing and machine learning. Given a trained RBM, the authors introduce a penalty-based denoising scheme that admits a simple QUBO form, for which they derive the statistically optimal penalty parameter as well as a practically-motivated robustness modification. The denoising step only needs to solve a QUBO admitting a bipartite graph representation, and so is well-suited for currently available quantum annealers. As QA has also shown promise for training BMs, this could yield a full image denoising model where both the model training and image denoisinghappen via QA. The denoising scheme was tested on a D-Wave Advantage device with good results, and larger problems were also successfully tackled through classical simulations. It's also worth noting that this approach is not limited to the QA framework, but is valid for any platform that supports a QUBO formulation. Moreover, it can be applied to the denoising of any binary data, not just images, making it a very general and widely applicable tool.

\subsubsection{Experimental Proposals and Architectural Designs for Quantum Annealing} 

One of the primary challenges in quantum annealing is transforming a given problem instance into an embbedded Ising model on a native hardware graph, such as the Chimera graph on early generations of D-Wave architectures. While the D-Wave machine can provide heuristic solutions to these Ising instances rapidly, it does not guarantee true optima. Conducting multiple runs may increase the chances of finding an optimal solution \cite{juenger2019performance}. 
 
Focusing on degree-bounded minimum spanning tree problems, experiments on the D-wave quantum annealer demonstrated strategic pausing within a specific annealing timeframe \cite{Gonzalez_Izquierdo_2021} significantly improves the probability of success and time-to-solution, extending the theoretical understanding of how embedding parameters interact with annealing parameters. 
 
However, a subsequent study revealed challenges in using quantum annealers for specific types of quantum simulations. Specifically, when probing thermal expectation values using the TFIM, the D-Wave 2000Q quantum annealer showed limitations in accurately reproducing these values \cite{Izquierdo_2021}. The study highlighted the dependency of the annealer's effectiveness on the quench rate and the susceptibility of different observables to this process. This suggests that while quantum annealers show promise in certain types of optimization problems, their ability to simulate specific quantum phenomena can vary significantly based on the nature of the problem and hardware constraints. 

Optimization problems requiring embedding, in particular graph coloring problems, were also considered in an extensive study of the different generations of D-Wave processors, from the D-Wave Two all the way to the Advantage~\cite{Pokharel_2023}, which demonstrated that the hardware upgrades and optimization of operational parameters---like anneal times and ferromagnetic couplings---that have been developed over the alst decade have made great strides towards improving quantum annealing performance.

Additionally, novel approaches to the graph minor-embedding problem, crucial for optimizing quantum annealing processes, have been explored. A monolithic integer programming approach and a decomposition approach, both hardware-agnostic, have been developed to provide more efficient alternatives to heuristic methods and identify infeasible instances \cite{bernal2019integer}. 
 
Moreover, recent advancements include the exploration of tailored quantum annealers for solving complex quantum models. In~\cite{Levy_2022}, the authors introduce a low-weight encoding to represent the Fermi-Hubbard model in terms of Pauli operators. This approach allows for the model to be expressed with a locality of at most three, facilitating the execution on quantum annealers. Numerical simulations have demonstrated this method's ability to reach near-ground state solutions for varying system sizes and its robustness against control noise. This development underscores the potential of tailored quantum annealers in simulating complex quantum systems, such as those found in materials physics, and represents a significant step forward in the field \cite{bernal2019integer}. 
 
Finally, the success in implementing algorithms for computing free energies on IBM’s quantum processor highlights the critical role of architectural design in quantum computing. Coupled with the development of efficient quantum circuits for dihedral gauge theories, as demonstrated in the Rigetti Aspen-9 quantum processor, these advancements showcase the ongoing progress in optimizing quantum computing architectures for a variety of sophisticated applications \cite{bassman2021computing}. 
 
\subsection{Combinatorial Optimization Problems} 
 
Advancements in combinatorial optimization problems within quantum computing are further enriched by innovative strategies in parameter setting for quantum algorithms. Formalizing concepts such as Perfect Homogeneity and the Classical Homogeneous Proxy for QAOA marks a significant step in this direction. A new heuristic for parameter optimization showcases the ability to effectively manage parameters up to 20 layers of QAOA \cite{sud2022parameter}. This advancement of QAOA to higher circuit depth was implemented through linear ramp schedules, inspired by quantum annealing, which simplify the parameter space and have been demonstrated to monotonically enhance approximation ratios as the number of QAOA layers. This approach paves the way for broader applications, potentially extending to varied problem classes and quantum algorithms beyond QAOA, marking a significant advancement in quantum computational capabilities \cite{sud2022parameter}. 
 
In exploring advancements in combinatorial optimization problems, significant strides have also been made in understanding the robustness of quantum algorithms under noise. A pivotal achievement is the development of an exact combinatorial expression for calculating the likelihood of maintaining particle number subspaces in the face of local depolarizing noise. This theoretical advancement has been crucially applied to benchmark the robustness of the XY variant of the Quantum Approximate Optimization Algorithm (XY-QAOA), specifically in addressing the Max-k-Colorable-Subgraph problem. The research underlines the profound influence of problem encoding on the algorithm's robustness, highlighting the necessity for innovative encoding methods and error mitigation strategies in noisy quantum optimization scenarios \cite{Streif_2021}. 
 
New insights from the study of QAOA with gradually changing unitaries reveal that the success of linear ramp schedules may extend well beyond shallow circuits. The discrete adiabatic theorem has been applied to understand the performance of QAOA at greater depths, particularly shedding light on the Ridge region of the QAOA performance diagrams. This region maintains algorithmic effectiveness with fewer parameters, challenging the notion that deeper circuits are always necessary for better performance \cite{kremenetski2023quantum}. 

Furthermore, the problem encoding's role in algorithm robustness has been underscored by studies like the XY-QAOA applied to the Max-k-Colorable-Subgraph problem demonstrating a correlation between encoding choices and noise resilience. These findings emphasize the importance of strategic problem encoding, especially as QAOA is pushed to higher depths where maintaining performance amidst errors is critical \cite{kremenetski2023quantum}. 
 
The efficacy of the proxy used to estimate the state and cost expectations of QAOA for smaller $\gamma$ values and for quantum annealing-inspired schedules affirms the relevance of these schedules for important parameter regimes. However, the research indicates that this proxy may not always hold for arbitrarily chosen parameters \cite{sud2022parameter}. 
 
Incorporating these new understandings, such as the changes in eigenstate connectivity highlighted in the recent literature, underscores the nuanced approach required in QAOA parameter optimization. It opens the door to further exploration into the absolute performance of QAOA on the Ridge and its scalability with larger problem sizes \cite{kremenetski2023quantum}. 

In optimizing quantum heuristics, the introduction of the Mixer-Phaser Ansätze represents a significant leap forward. This approach, which allows for the efficient compilation of circuits, has been empirically matched in performance with QAOA using less than half the circuit depth on most superconducting qubit processors \cite{LaRose_2022}. 
Ref. \cite{LaRose_2022} presents a variation that merges the spirit of Variational Quantum Eigensolvers with the QAOA's advanced mixers, guided by cost function-derived operators. The numerical results support the potential of the mixer-phaser ansatz, named QAMPA, for the next generation of NISQ era quantum algorithms. The authors suggest QAMPA as a compilation-advantageous method for hard-constrained combinatorial optimization problems, positing further research towards deploying QAMPA solvers on quantum hardware and assessing their comparative advantage in real-world scenarios \cite{LaRose_2022}. 
 
By encompassing the latest findings on problem encoding, parameter optimization, and the discrete adiabatic theorem's applications, the field is moving towards a more sophisticated and efficient use of QAOA, even as circuit depths increase. 

A different realization of a physical system whose dynamics can converge to the low-energy configuration of disordered Ising spin models is the (optically) Coherent Ising Machine (CIM), whose (noiseless) behavior can be represented by interconnected sets of regular differential equations (ODEs). In~\cite{taassob2023}, the authors consider the possibility of simulating these systems through Neural Operators, a novel approach in which a neural network can be trained to approximate a functional. While neural networks have been used for decades to solve combinatorial optimization problems, Neural Operators are a recent development which have shown great promise in solving inverse problems and to replace multi-physics simulations of PDEs with a less time and energy-intensive data-driven approach.

They investigate the proof-of-concept question of whether we can use these neural operator methods to simulate the interconnected sets of ODEs by applying them to a set of spin-glass Ising problems of the Sherrington-Kirckpatrick type, and benchmarking their performance for simulating of the continuous dynamics, as well as that of the resulting trained network as an Ising solver, i.w. verifying whether the postprocessed bitstrings after the neural inference are of quality as solutions to the Ising problem. Deep Operator Networks (DeepONet) were used for the neural architecture,  which appear scalable and easily deployable to simulate coupled ODEs.

After training, the network proved capable of efficiently optimizing problems and initial conditions up to and beyond 175 spins, which is a highly non-trivial task, especially in a regime with many coupled equations undergoing semi-chaotic dynamics. While these results were limited by a rather small computational cut-off for training, they are promising as it is expected that they could generalize to higher sizes if sufficient data is allowed.
 
\subsection{Quantum State Preparation and Analysis}
\subsubsection{Quantum Many-Body Scar (QMBS) States}
 
Quantum many-body scar states, known for their unique entanglement and correlation properties, play a pivotal role in sustaining long-lived coherent dynamics. The preparation of the superposition state $|\xi\rangle$, a ground state candidate for adiabatic state preparation, is explored \cite{gustafson2023preparing}. This state, characterized by periodic dynamics and modest entanglement scaling, presents a feasible target for state preparation on digital quantum computers, despite challenges like Trotter errors. 
 
Ref. \cite{gustafson2023preparing} delves into the lifetime of dynamics exhibited by these scarred states. It highlights that while some scarred states can be prepared trivially, others, especially those with area-law entangled superpositions, require more complex methods. Two approaches are considered: a linear-depth circuit for perfect fidelity preparation and a probabilistic method with a tradeoff between circuit depth and postselection success probability. This flexibility is advantageous for implementation on near-term quantum hardware. 
 
The research acknowledges the limitations of Near-Term Intermediate-Scale Quantum (NISQ) devices, quantifying potential errors using the Bhattacharyya distance. Experiments on Rigetti's Aspen Quantum Processing Units (QPUs) demonstrate a balance between error and post-selection success probability, influenced by the size of circuit fragments and block size $m$. Various error mitigation techniques, including iterative Bayesian unfolding and randomized compilation, are applied to improve accuracy and manage sample variance \cite{gustafson2023preparing}. 
 
The preparation and analysis of quantum many-body scar states are significant for probing non-ergodic quantum dynamics and understanding how certain states evade thermal equilibrium. Despite the inherent limitations of NISQ devices, the successful preparation and analysis of these intricate states highlight their potential for deeper experimental exploration in quantum computing. 
 
\subsubsection{Parameter Initialization Methods for Quantum Circuit Simulations}
 
The initialization of parameters in quantum circuits, especially for the unitary coupled cluster singles and doubles (UCCSD) ansatz, is critical in quantum state preparation.  Two classical methods, the second-order Moller-Plesset perturbation theory (MP2) and the coupled cluster method with singles and doubles (CCSD), are explored for generating starting parameters. To facilitate these simulations, a technique is employed that simplifies quantum simulations by discarding less critical components of the molecular system, particularly useful when handling extensive systems \cite{hirsbrunner2023mp2}. 
 
Comparisons between these methods show that initializing UCCSD simulations with CCSD parameters generally leads to more accurate results than MP2, with energy values closer to expected benchmarks across a range of molecules. This trend is consistent for almost all tested cases, except for a specific molecule in a stretched configuration, suggesting CCSD parameters as the preferable choice for setting up quantum chemistry simulations in the current quantum computing landscape. As quantum computers continue to evolve and grow, yet still remain within the bounds of classical simulation capabilities, the use of CCSD parameters stands out as a more effective approach for initializing these complex quantum simulations \cite{hirsbrunner2023mp2}. 

Parameter initialization also plays an important role in Variational Quantum Algorithms (VQAs), which circumvent the quick decoherence and limited number of qubits available in current quantum computing devices by using a classical procedure to select gate parameters (namely rotation angles) for a quantum circuit representing a problem of interest. When the circuit is evaluated on quantum hardware, and the final states obtained through a large number of shots will follow a distribution representing the solution to the computational problem. Then follows a feedback loop where a classical optimization algorithm selects the parameters for the quantum circuit based on a measure of the output bitstrings to find the optimal parameters for a quantum circuit using the variational principle.

One of the challenges with the classical part of VQAs is that it corresponds to a black-box optimization problem that is generally non-convex, requiring global optimization strategies. There is a trade-off between cost and accuracy; a high number of runs on the quantum computer (which is costly) is needed to measure the circuit accurately. One possible solution is using basic Bayesian optimization (BO) methods to globally optimize quantum circuit parameters, which has shown good potential. These are a family of sample-efficient zeroth-order optimizers which has proven successful at solving VQAs (among other problems). BO’s sample efficiency results from using observations from the quantum circuit to construct a statistical surrogate model known as a Gaussian Process (GP), which generalizes a multivariate normal distribution to function space. The GP’s ability to quantify the model uncertainty allows to systematically trade off between exploration and exploitation of the parameter space.

In~\cite{Sorourifar2024}, the authors propose two modifications to the basic BO framework to provide a shot-efficient optimization strategy for VQAs. They show that a significant increase in performance can be achieved by encoding priors into the GP kernel function and surrogate model. The kernel prior endows the GP with knowledge of the parameter’s periodicity, which they find helpful in the limited circuit observation regime. At the same time, the topological prior provides a better starting model by utilizing large quantities of low-shot circuit measurements. 

\subsubsection{Population Transfer in Quantum Random Energy Model} 
 
Exploring further into quantum state preparation, a study on the population transfer protocol within the Quantum Random Energy Model offers intriguing insights. This research focuses on the energy matching problem, aiming to find multiple approximate solutions to combinatorial optimization problems. A key aspect of the study is observing how the population transfer protocol influences the delocalization process, assessed by measuring the Shannon entropy of the time-evolved wavefunction. This approach helps identify various dynamical phases of the model and assesses the effectiveness and uniformity of the population transfer. Particularly, the study finds that the protocol is most effective when the transverse-field parameter is close to the critical point of the Anderson transition.  While no strong speedup compared to random search is observed at accessible system sizes, the insights gained are valuable for understanding quantum state preparation and analysis in the context of complex quantum dynamics and optimization problems \cite{parolini2020multifractal}.  
 
A significant advancement in quantum state preparation and analysis is the development of the Two-Unitary Decomposition (TUD) algorithm, which addresses the challenge of simulating open quantum systems. Open quantum systems are rarely isolated and are often influenced by their environments, leading to non-unitary open quantum evolution. This evolution is fundamental to understanding dissipation, decoherence in quantum systems, and a variety of phenomena including thermalization and transport in strongly correlated systems \cite{Suri_2023}. 
 
The TUD algorithm provides a novel method to decompose any d-dimensional contraction operator with non-zero singular values into two unitaries. This approach, which uses the quantum singular value transformation (QSVT) algorithm, avoids the classical overhead associated with singular value decomposition (SVD). It requires only a single call to the state preparation oracle for each unitary and can significantly reduce calls to the encoding oracle \cite{Suri_2023}. This method has potential applications not only in simulating open quantum systems but also in linear algebra and quantum machine learning. 
 
The TUD algorithm represents a state-of-the-art technique for simulating open quantum systems. It provides a more efficient way of expressing non-unitary Kraus operators as a sum of two unitaries, which can be deterministically implemented. This approach is essential for advancing our understanding of complex systems, particularly those embedded in non-trivial environments like biological and chemical systems. The ability to simulate these systems fully can potentially reveal new physics and mechanisms that have been inaccessible with conventional techniques \cite{Suri_2023}. 
 
\subsubsection{Efficient Hamiltonian Simulation, Compression, and Thermal State Generation}
 
A significant advancement in Hamiltonian simulation for quantum computers introduces the Self-consistent Quantum Iteratively Sparsified Hamiltonian (SQuISH) method. This approach, crucial for electronic structure problems, addresses the challenge posed by the computational scaling of Hamiltonian terms with the number of orbitals, which typically grows as $N^4$ \cite{chamaki2022selfconsistent}. SQuISH employs a novel strategy of aggressively truncating Hamiltonian terms while maintaining targeted accuracy, particularly beneficial for NISQ applications where quantum hardware limitations are a key concern. 

The core innovation in SQuISH is the use of an iteratively updated ground state wavefunction to determine the relative importance of each Hamiltonian term \cite{chamaki2022selfconsistent}. This iterative process involves finding the ground state wavefunction and energy of a truncated Hamiltonian and checking for convergence in successive iterations \cite{chamaki2022selfconsistent}. SQuISH has shown efficacy in handling electronic structure problems, and its generic nature suggests potential benefits in additional applications where Hamiltonian size is a computational bottleneck. 

In parallel, advancement in quantum machine learning and quantum simulation with the introduction of Quantum Hamiltonian-Based Models (QHBMs) and the Variational Quantum Thermalizer (VQT) has been made. These models represent a paradigm shift in quantum-probabilistic hybrid variational learning, efficiently decomposing tasks to learn both classical and quantum correlations. QHBMs are particularly adept at generating thermal states of a given Hamiltonian and target temperature, making them well-suited for tasks requiring the simulation of thermal and mixed quantum states \cite{verdon2019quantum}. 

\subsection{Advanced Quantum Algorithms and Applications} 

\subsubsection{Quantum-Enhanced Deep Learning Models}
 
Ref. \cite{Templin2023}  explores the development of unsupervised Variational Autoencoder (VAE) models with discrete latent variables for anomaly detection in commercial aeronautics data. It focuses on two types of models: a Bernoulli model using a factorized Bernoulli distribution as prior, and a more flexible and expressive Restricted Boltzmann Machine (RBM) model integrated into the VAE's latent space. The RBM model, in particular, demonstrates compatibility with quantum computing, as its negative phase states can potentially be derived from quantum Boltzmann sampling. This feature enables the RBM model to perform on par with its Gaussian counterpart in anomaly detection tasks, highlighting the promise of quantum-enhanced machine learning models in handling complex, multifactorial data.  
 
\subsubsection{Hybrid Quantum-Classical Frameworks}
 
A groundbreaking approach introduced in Ref. \cite{brown2023copositive} utilizes Ising solvers within a hybrid quantum-classical framework for the global optimization of mixed-binary quadratic programs (MBQP). The novelty of this method lies in its use of a convex copositive reformulation of MBQPs, solved via a cutting-plane algorithm where the classical computation is polynomial in time. This technique effectively shifts the complexity of NP-hard problems onto the Ising solver, potentially harnessing quantum accelerators' capabilities without requiring an exhaustive search by classical algorithms. 

The framework's classical components scale polynomially with the number of constraints, suggesting that quantum speed-ups, even if not explicitly characterized, can be exploited. This aligns with the advancements in parameter optimization for QAOA, which seek to push the boundaries of quantum computing efficiencies. The integration of such copositive programming with Ising solvers could represent a significant step forward in solving complex optimization problems, leveraging the strengths of both quantum and classical algorithms \cite{brown2023copositive}. 

Another promising approach combining quantum data and optimization heuristics with classical machine learning was introduced in~\cite{Khan2024} to tackle molecular dynamics simulations. Performing these simulations on quantum computers would be a natural choice and lead to exact results, but is currently limited due to noisy hardware, the costs of computing gradients, and the number of qubits required to simulate large systems. As this paper demonstrates, these issues can be mitigated using some recently developed machine learning techniques. In particular, the authors propose a new method to make efficient use of a small number of high accuracy VQE energies, combining transfer learning with techniques for building machine-learned potential energy surfaces. The potential energy surface is represented as a neural network that takes the coordinates of the system as an input and outputs an energy, and can be efficiently evalluated and differentiated on a classical processor. 

To reduce the quantum resources needed, the model is initially trained with data derived from low-cost techniques, such as Density Functional Theory, then refined and corrected using Unitary Coupled Cluster data. Once successfully trained, the model produces the energy gradient predictions necessary for dynamics simulations, which cannot be readily obtained from quantum hardware alone. This is showcased in the paper by accurately modeling the internal dynamics of the water monomer and both the inter- and intramolecular dynamics of the water dimer.

While the VQE is a promising and powerful choice for the NISQ era with a multitude of applications, it suffers from the detrimental effects of barren plateaus, noisy hardware and slow convergence, in particular when optimizing from random starting parameters. In~\cite{Khan2023}, the authors propose the variational tensor network eigensolver (VTNE), an approach that mitigates these issues by classically pre-optimizing circuits. It works by classically simulating VQE by approximating the parameterized quantum circuit (PQC) as a matrix product state (MPS) with a limited bond dimension, which determines how much work is saved on quantum hardware. Through benchmarking on Fermi-Hubbard models, the authors showcase the effectiveness of using an approximate tensor network backend for VQE, leading to accurate ground energy estimation for the 1D case and efficient circuit initialization for 2D. This approachs opens the door to extending the applicability and scalability of VQE on near-term quantum hardware.

Another task that VQE can be applied to is that of gound state preparation for quantum systems on quantum hardware. This has a wide range of uses for condensed matter, quantum chemistry, and many other fields. While fault-tolerant hardware is still in development, it is critical that short-depth efficient circuits for state preparation are available. The VQE is a good candidate for moderately sized systems to efficiently construct the states where the circuits are optimized on the quantum hardware. However, its potential has not yet been fully realized, partly due to the need for optimization in the presence of noise.

One general approach to this problem is to forego optimization performed on the quantum hardware in favor of heuristic classical computational approaches to optimize the parameters of a quantum circuit before running on actual hardware. This still leaves the question of how to go beyond the classically optimized circuit effectively to find quantum advantage. A specific way of doing this is presented in~\cite{Gustafson2024}. The authors draw inspiration from a geometry optimization algorithm developed for noisy electronic structure calculations, where a surrogate model (density functional theory) is used to calculate a Hessian for the density functional potential energy surface, which is expected to be near the exact minimum geometry. 
is amenable to the 

This algorithm is adapted to the optimization of variational quantum eigensolver circuits in chemistry and condensed matter by using an approximate (classical CPU/GPU) state vector simulator as a surrogate model, through which an approximate Hessian is efficiently calculated and passed as an input for a quantum processing unit or exact circuit simulator. 

The effectiveness of this method is demonstrated through application for a number of molecules and quantum spin models, including using an IBM quantum computer for the transverse Ising model using 40 qubits. The new optimizer outperforms standard methods in the presence of sampling noise, and will lend itself well to parallelization across quantum processing units. 

\subsubsection{Quantum Hamiltonian Based Models and Variational Quantum Thermalization}
 
SQuISH has been extended to include multi-reference and non-iterative truncation methods. The multi-reference version, called "multi-reference SQuISH," considers both ground and excited state energetic contributions, useful in dynamics and creating a selected energy space for truncation processes \cite{chamaki2022selfconsistent}. Furthermore, SQuISH can be combined with algorithms like the Variational Quantum Eigensolver (VQE) for non-iterative truncation. This allows the use of a sufficiently accurate wavefunction, obtained via VQE, to rank and truncate Hamiltonian terms in a single iteration, offering a practical approach for Hamiltonian compression \cite{chamaki2022selfconsistent}. 

In addition to these truncation methods, the development of Quantum Hamiltonian Based Models (QHBMs) represents a significant advancement in the field of quantum machine learning and quantum simulation \cite{verdon2019quantum}. QHBMs, particularly adept at generative modeling, focus on two tasks: Quantum Modular Hamiltonian Learning (QMHL) and Variational Quantum Thermalization (VQT). QMHL involves learning the modular Hamiltonian of a target mixed state for its reproduction, while VQT is geared towards generating approximate thermal states of a known Hamiltonian  \cite{verdon2019quantum}. These models embody a novel approach in quantum-probabilistic hybrid variational learning, efficiently decomposing tasks to learn both classical and quantum correlations, and thus expanding the utility of quantum simulations. 

\subsection{Software Development} 
\subsubsection{Software Tools \& Development}

TensorFlow Quantum remains a key platform for quantum machine learning models and algorithms \cite{broughton2021tensorflow}. It supports a range of advanced quantum learning tasks, including meta-learning, layerwise learning, Hamiltonian learning, sampling thermal states, variational quantum eigensolvers, classification of quantum phase transitions, generative adversarial networks, and reinforcement learning \cite{broughton2021tensorflow}. 
 
Quantum circuit simulation is crucial for both demonstrating quantum supremacy and verifying the performance of quantum hardware. The primary proposal for achieving quantum supremacy with NISQ devices involves sampling bit-strings from a random quantum circuit (RQC), a task believed to be classically infeasible for large enough systems. The Flexible Quantum Circuit Simulator (qFlex) is introduced as a tool for simulating RQCs efficiently, providing a new benchmark for quantum supremacy experiments and allowing for improved verification of quantum hardware performance\cite{Villalonga2019}. 

In addition to quantum algorithms, there are also several quantum-inspired approaches that have been developed as a means for benchmarking against quantum algorithms, but also as powerful hardware algorithms in their own right \cite{chowdhury2025pushingboundaryquantumadvantage}. In several projects within the Feynman quantum academy these quantum-inspired techniques \cite{Aadit_Nonlocal_Monte_Carlor_2023}, \cite{Aadit_APT-solver_An_Adaptive_2023} and benchmarking tools \cite{Bernal_Neira_Stochastic_Benchmark_toolkit_2023, Bernal2024}
 have been used for optimization and machine learning based projects. By open-sourcing these codes, interns gain valuable experience developing and documenting reusable code so that they can continue working with collaborators  to use and advance the techniques in future projects.

\subsection{Benchmarking quantum algorithms and systems} 
\subsubsection{Enhanced Techniques for Quantum Benchmarking}

Quantum gate fidelity benchmarking is essential for evaluating quantum algorithms' performance. While quantum state tomography is a traditional approach, it can be confounded by state preparation and measurement (SPAM) errors. To circumvent this, the Generalized Character Randomized Benchmarking (GCRB) method has been developed, utilizing the algebraic properties of unitary groups, denoted as \( G \), which dictate quantum transformations \cite{Claes_2021}. Expanding Randomized Benchmarking to include non-multiplicity-free groups allows for more accurate gate fidelity estimates and applies to a broader range of gate groups. It includes subspace Randomized Benchmarking for gates symmetric under SWAP, a new leakage RB protocol, and a scalable protocol for the matchgate group, one of the few non-Clifford scalable RB protocols. GCRB leverages the characters of these representations to accurately estimate the average fidelity of a set of quantum gates, bypassing SPAM errors. Its efficacy has been demonstrated with the Clifford group, integral to quantum error correction algorithms, showcasing its utility even in the presence of complex, non-multiplicity-free structures \cite{Claes_2021}. 

Recently, there has been a significant advancement in benchmarking quantum algorithms that conserve particle number, particularly those susceptible to noise. A symmetry-aware error mitigation scheme has been introduced, which is less resource-intensive than conventional quantum error correction. This scheme offers a targeted approach for correcting errors in quantum circuits that preserve specific symmetries, such as particle number symmetries \cite{Streif_2021}. The efficacy of this approach has not only highlighted the importance of novel encoding techniques but also opened new pathways for NISQ algorithms that maintain general symmetries, which are crucial as we progress toward more robust quantum computing. 

A novel technique for estimating error-mitigated expectation values on noisy quantum computers has been developed. This method utilizes shadow tomography on a logical state, enabling a memory-efficient classical reconstruction of the noisy density matrix. It marks a significant reduction in quantum and classical resources overhead compared to existing methods like subspace expansion and virtual distillation \cite{hu2022logical}. 

\subsubsection{Quantum Performance Benchmarks in Optimization Algorithms}
 
Ref. \cite{lubinski2024} introduces a framework for benchmarking quantum performance in optimization tasks using Quantum Annealing (QA) and Quantum Approximate Optimization Algorithm (QAOA). This approach is structured to provide insights into unique aspects of quantum computing while being recognizable to optimization practitioners. The methodology exercises multiple components of integrated hybrid quantum-classical systems and captures, analyzes, and presents performance metrics uniformly across quantum computing architectures. 
 
This benchmarking approach is particularly applied to the Max-Cut problem, an NP-HARD problem representative of a class of optimization challenges that are simple to specify yet difficult to solve. It forms part of the open-source QED-C Application-Oriented Benchmark suite, enhanced with specific quality and temporal metrics. Applicable for both gate model and quantum annealing computers, it illustrates the trade-off between resource usage and solution quality. 
 
Both the QAOA and QA algorithms converge to solutions through iterative processes, with the quality of results being influenced by the execution time allowed for these algorithms. The financial implications of quantum computation are often tied to the cumulative quantum execution time. A notable development is the use of containerized execution environments, like the Qiskit Runtime service, which significantly reduce total elapsed execution time. This approach prevents queue times from accumulating for each quantum ansatz execution in a hybrid program, offering benefits for users without privileged access to quantum hardware.  
 
Execution times using parameterized circuits on systems like the IBM Quantum Guadalupe have shown that eliminating circuit transpilation before each execution can reduce total elapsed time significantly. Execution on Quantum Annealing Computers, as shown in the benchmark results, includes cumulative elapsed time and quantum execution time, with the cost of using the QA system being tied to the quantum execution time. 

More recently, a very general framework was introduced to benchmark the performance of hybrid quantum-classical algorithms and physics-based algorithms, based on a characterization of parameterized stochastic optimization solvers~\cite{Bernal2024, Bernal_Neira_Stochastic_Benchmark_toolkit_2023}. It is well suited for stochastic optimization methods, which include quantum methods for optimization such as quantum annealing and gate-based variational parametric algorithms, and provides a more holistic reporting on algorithmic performance, by allowing the comparison of different setups for a given solver, which is useful for parameter setting and tuning tasks, and laying out general rules for a benchmarking procedure to be objective.

Building on the benchmarking frameworks,~\cite{a15100356} introduces several tailored ansätze for solving the combinational circuit fault diagnostic (CCFD) problem. These ansätze, which are more closely aligned with the structure of the underlying optimization problems, demonstrate better performance compared to more generic approaches. The results from this study support the notion that ansätze exploiting the problem structure can significantly improve performance in optimization tasks like CCFD problems. This finding is particularly relevant for many NP-hard combinatorial optimization problems, indicating a promising direction for future research in quantum algorithms and QAOA-based solutions. 

\subsection{Quantum-compatible Machine Learning Algorithms}

Over the past few years, there have been significant developments in machine learning methods, both in discriminative modeling such as classification and generative modeling to support analysis based on an understanding of the probability distribution of data. One of the key areas of interest within the QuAIL team is understanding how to integrate and leverage quantum probability distributions and/or samples from quantum processors in these powerful machine learning models. Investigations into how to effectively leverage samples from quantum processors into state-of-the-art machine learning models, such as VAE, GAN, Invertible flows, U-Nets and also consider the multiple challenges, involving latency, connectivity, and noise of near-term quantum processors have been the focus of the team. 

This section delves into these hybrid machine learning models, examining how quantum samples, especially from Boltzmann or Born machines, are integrated within the latent spaces of these models. In several cases, these quantum-compatible methods based on discrete latent space models were shown to provide value over the baseline standard approach, demonstrating continued opportunity for quantum-inspired machine learning approaches. However, demonstrating use of samples from noisy, sparse latent distributions representative of modern quantum hardware is still a challenge.

The union of Quantum Computing with machine learning can be carried out in four ways:
\begin{enumerate}
    \item The Classical-Classical Approach (CC), seen in the use of quantum-inspired methods in discrete latent space models, showcases the potential for quantum influence in enhancing traditional algorithms.
    \item The Classical-Quantum Approach (CQ) is exemplified by models like the Quantum-assisted Variational Autoencoder (QVAE), where quantum computing elements are utilized to process and interpret classical data.
    \item The Quantum-Classical (QC) paradigm, observed in models optimizing quantum heuristics with classical machine learning techniques, reflects the application of traditional methods to data generated by quantum processors.
    \item Lastly, the Quantum-Quantum (QQ) approach, though not explored in this section, represents the full fusion of quantum data and quantum computational methods.
\end{enumerate}

\subsubsection{Quantum-assisted Variational Autoencoder}
In their groundbreaking work, N. Gao et al. introduced a novel approach in Quantum Machine Learning (QML) by utilizing a quantum loss function within the framework of a Variational Autoencoder (VAE) \cite{Gao2020}. This quantum-enhanced VAE adeptly handles complex probability distributions in high-dimensional data spaces, addressing key challenges of scaling classical machine learning algorithms to such intricate datasets. 

The core of this research lay in the Quantum-assisted VAE (QVAE), which elegantly integrates samples from a quantum annealer into the prior of a discrete VAE, enabling the modeling of powerful and flexible distributions with the potential to overcome limitations faced by classical counterparts. The encoder comprised two fully-connected layers leading into a hierarchical posterior, allowing for more complex posterior distributions and thus optimize a tighter ELBO. To avoid overfitting, the decoder network is simpler by only featuring two full-connected layers. 

Further enhancing the capabilities of the QVAE, N. Gao et al. utilized the tunable quantum effects associated with Quantum Boltzmann Machines (QBM), particularly the transverse field. This field affected the expressiveness of the prior and allowed for narrowing the latent distribution while maintaining the network’s ability to learn the occupancy of various states.

An integral part of the study involved comparing the performance of a Restricted Boltzmann Machine (RBM), a classically simulated QBM, and a QBM trained with samples drawn from a quantum annealer. The research team utilized the D-Wave 2000Q quantum annealer for this purpose, a device featuring 2048 qubits each of degree 6. Due to hardware limitations, any RBM with more than 6 nodes per side had to be embedded using a chain coupling strength of -1, the maximum allowed by the D-Wave 2000Q. At the end of an anneal, the value of a logical variable was determined by a majority vote, adding a layer of robustness to the modeling process.

Post-training, the QVAE's encoder was utilized to index the dataset efficiently without the need for additional discretization, thanks to the binary nature of its latent space. The encoding algorithm mapped similar objects to the same bit strings within the latent space. In terms of the search process, an inverted index (hash map) was constructed to map a bit string to all data points that have been mapped to that bitstring. Since every item is only
processed once and independently this operation, the original data can be removed from main memory after processing reducing memory usage. For queries, the QVAE first encoded the item and then sorted all occupied bit strings by their Hamming distance to the query embedding. The iterative comparison of items based on their Hamming distance offered a fine balance between reducing the memory footprint and enhancing the resolution of the search space.

The results were compelling:

\begin{itemize}
\item \textit{Embedded Proximity and Hamming Distance:} Experiments showed that in the compressed space, Hamming distance effectively approximated Euclidean distance in the original space, as validated through k-ANNS on the MODIS dataset. This underscores the efficiency and accuracy of quantum-assisted techniques in high-dimensional data retrieval.
\item \textit{Impact of the Transverse Field:} Adjusting the transverse field parameter influenced the distribution's characteristics and search speed, with optimal speedup observed at certain ranges. This highlights the precise control quantum models offer in data representation and optimization for high-dimensional data.
\item \textit{Memory Consumption:} The quantum model demonstrated superior memory efficiency over methods like HNSW and LSH, particularly in managing large datasets such as the complete MODIS dataset. This efficiency emphasizes the quantum model's practicality and scalability for large-scale, high-dimensional data tasks.
\end{itemize}

\subsubsection{Deep Generative Models}
\cite{Templin2023} investigated the capabilities of unsupervised deep generative models, particularly focusing on VAEs) with discrete latent variables (DVAE). The study presented two distinct DVAE models: one with a factorized Bernoulli prior and another that integrated a Restricted Boltzmann Machine (RBM) as its prior. In the model design, the RBM DVAE model is highlighted for its flexible and expressive nature, which incorporates an RBM into the VAE's latent space. This model is especially noted for its compatibility with quantum computing techniques, as the RBM's negative phase states can be derived from quantum Boltzmann sampling, offering a unique approach to handle complex, multifactorial data \cite{Templin2023,Wilson2021}.

The paper discusses the specific design elements and training strategies for these models, including the use of the Gumbel-softmax trick for differentiability in the discrete models and the introduction of a hyperparameter $\beta$ in the ELBO objective function to optimize anomaly detection performance \cite{Templin2023}.

Performance evaluation of these models is a key aspect of the study. The researchers utilize metrics like precision, recall, and F1 score to assess anomaly detection capabilities in flight operations data. The results demonstrate that the RBM DVAE model performs on par with the Gaussian model and outperforms the Bernoulli model, showcasing the effectiveness of a discrete deep generative model against its Gaussian counterpart in anomaly detection tasks \cite{Templin2023}.

Furthermore, the research discusses the robustness of the RBM model in varying anomaly types and flight phases, as well as its potential integration with quantum computing for enhanced performance. The study emphasizes the universal applicability of these models beyond aeronautics data, highlighting their potential in other time-series data applications \cite{Templin2023}.

\subsubsection{Quantum-assisted associative adversarial networks}
In exploring the integration of quantum annealing with deep learning architectures, Ref. \cite{Wilson2021} introduced the Quantum-assisted Associative Adversarial Network (QAAAN). This model incorporates quantum annealing to train a Boltzmann Machine (BM) that optimizes the feature distribution extracted by the discriminator network in the adversarial model. The unique capabilities of quantum computing are leveraged to find more effective representations in the latent space, thereby enhancing the generation of realistic data. This approach, particularly significant for its exploration of reparametrization of discrete variables, demonstrates a crucial step in integrating quantum models with traditional neural networks \cite{Wilson2021}.

In the QAAAN study, the researchers investigated various topologies for the probabilistic graphical models in the latent space, including complete, symmetric bipartite, and Chimera topologies.  It's noteworthy that the choice of topology affects both the model's learning rate and quality, similar to how transverse fields influenced the quantum variational autoencoder’s performance in Ref. \cite{Wilson2021b}. This insight is crucial for future developments in quantum-enhanced deep learning, as it highlights the need for careful consideration of model architecture to harness the full potential of quantum computing.

The QAAAN's exploration of topological variations extends to practical performance metrics. The researchers utilized the Inception Score and the Frechet Inception Distance to assess the model's ability to generate realistic data. This evaluation was pivotal in establishing the QAAAN's efficacy in real-world applications. The research also demonstrated the model's scalability by successfully applying it to complex datasets, such as the LSUN bedrooms dataset. This scalability indicates the potential for quantum-assisted models in handling large-scale, high-dimensional data challenges \cite{Wilson2021}.

\subsubsection{Quantum-compatible Probabilistic Wildfire Segmentation}
In Ref. \cite{Asanjan2023}, an innovative approach in wildfire detection using a supervised deep generative model is introduced. This model, employing VAE techniques for processing satellite imagery, is adept at enhancing wildfire segmentation accuracy. The significant aspect of this approach is its ability to handle high-dimensional datasets, resonating with the quantum-assisted VAE approach by N. Gao et al. \cite{Gao2020} in dealing with complex probability distributions.  The Probabilistic U-Net model developed in this study is notable for its stochastic modeling, generating diverse segmentations and acknowledging the inherent uncertainties in wildfire detection. This aligns with quantum-inspired methods that incorporate stochasticity and uncertainty in model predictions \cite{Asanjan2023}.

Key to this model is its "what-if" scenario exploration, simulating different wildfire outcomes based on changing NDVI dynamics. This feature is vital for understanding the physical relationships in wildfire spread, a significant advancement in quantum-compatible machine learning. The model's superior accuracy and flexibility in wildfire detection and segmentation were highlighted through a comparative analysis against baseline stochastic models. It demonstrated a more comprehensive understanding of the physical interplay between NDVI and wildfire, showcasing its robustness in various wildfire situations. This research reflects the growing relevance of quantum-enhanced machine learning models in complex, dynamic systems like wildfire management, indicating the potential of quantum-inspired methods in environmental modeling and decision-making \cite{Asanjan2023}.

\subsubsection{Meta-learning}
VQAs, categorized under quantum heuristics, show great potential for practical quantum computing applications. The optimization of these algorithms for effective hardware performance is a critical area of focus. Ref. \cite{Wilson2021b} assesses the efficacy of a Long Short Term Memory (LSTM) recurrent neural network model (the meta-learner) in optimizing two quantum heuristics, comparing its performance traditional optimizers (Bayesian optimization, evolutionary strategies, L-BFGS-B and Nelder-Mead). 

The meta-learner outperforms traditional optimizers, such as Bayesian optimization, evolutionary strategies, L-BFGS-B, and Nelder-Mead, demonstrated superior performance in noisy environments. For example, in Fermi-Hubbard model problems, L-BFGS-B performance reduces by 0.35 whereas the meta-learner only reduces by 0.2, from around the same starting point \cite{Wilson2021b}. This is a strong indicator that meta-learning will be especially useful in noisy near-term quantum heuristics implemented on hardware. In addition to robustness, the meta-learner showed a higher frequency of reaching near-optimal solutions than the next best optimizer (evolutionary strategies) in noisy simulations \cite{Wilson2021b}. 

In addition to robustness, the meta-learner showed a higher frequency of reaching near-optimal solutions than the next best optimizer (evolutionary strategies) in noisy simulations \cite{Wilson2021b}. 

Looking ahead, the continued improvement of meta-learning methods is anticipated. Despite the current lack of investigation into their performance scaling to larger problem sizes, largely due to the challenges in simulating large quantum systems, the potential for these methods on hardware implementations is significant. The meta-learner introduced by \cite{Wilson2021b} applies a single model across various parameters (a 'coordinatewise' approach). Envisioning a 'qubitwise' approach, where distinct models are trained for each qubit's parameters in a given hardware graph, could open up new optimization avenues. Such a method might account for the unique physical characteristics of each qubit, potentially leading to more finely-tuned optimizations tailored to specific hardware environments. This direction underscores the importance of developing meta-learning methods that not only adapt to the complexities of quantum problems but also leverage the peculiarities of quantum hardware.

Ref. \cite{Wilson2021b} also notes the different implementations of QAOA used for Graph Bisection and MAX-2-SAT. Understanding the impact of mixer and initial state variations on performance, as well as characterizing the relative power of different QAOA mixers, remains an open area of research.

\subsubsection{RBM-Flow and D-Flow}

In \cite{oconnor2021}, we introduce EBM-Flows, which are Invertible Flow (IF) models with Energy Based Models (EBMs) as a trainable base distribution. Non-autoregressive IFs are a promising class of models that allow for exact likelihood training, unlike VAEs; can be efficiently sampled from, unlike autoregressive models; and do not require a discrimator network, unlike generative adversarial networks (GANs). In particular, we used a Restricted Boltzmann machine (RBM) as the EBM and introduced an EBM-Flow sub-class called RBM-Flow. Additionally, we introduced D-Flow, which is obtained from RBM-Flow by setting to zero all the coupling of the underlying latent RBM, and has the benefit that global features are meaningfully encoded as discrete labels in the latent space.

\end{document}